\documentclass[12pt,a4paper]{article}
\usepackage[margin={1.5cm,3cm}]{geometry}

\usepackage{amsmath,amssymb,cancel}
\usepackage{stackrel}
\usepackage{graphicx,hyperref}
\usepackage{epstopdf}
\usepackage{xcolor}

\usepackage[T1]{fontenc}
\usepackage[utf8]{inputenc}
\usepackage{authblk}

\newcommand{\beq}{\begin{eqnarray}}
\newcommand{\eeq}{\end{eqnarray}}

\newcommand{\nn}{\nonumber}

\usepackage[font=small,labelfont=bf]{caption}

\numberwithin{equation}{section}

\begin{document}
\renewcommand\Affilfont{\itshape\footnotesize}
\title{Searching for a $C$-function on the three-dimensional sphere}
\author[1]{C. G. Beneventano\thanks{\,\href{mailto:gabriela@fisica.unlp.edu.ar}{gabriela@fisica.unlp.edu.ar}}}
\author[2,3]{I. Cavero-Pel\'aez\thanks{\,\href{mailto:cavero@unizar.es}{cavero@unizar.es}}}
\author[1]{D. D'Ascanio\thanks{\,\href{mailto:dascanio@fisica.unlp.edu.ar}{dascanio@fisica.unlp.edu.ar}}}
\author[1]{E. M. Santangelo\thanks{\,\href{mailto:mariel@fisica.unlp.edu.ar}{mariel@fisica.unlp.edu.ar}}}
\affil[1]{Departamento de F\'isica, Universidad Nacional de La Plata, \protect\\  Instituto de F\'isica La Plata, CONICET-Universidad Nacional de La Plata, \protect\\ C.C.67 (1900) La Plata, Argentina}
\affil[2]{Centro Universitario de la Defensa, CUD, E-50009 Zaragoza, Spain}
\affil[3]{Departamento de F\'isica Te\'orica, Universidad de Zaragoza\\
E-50009 Zaragoza, Spain}
\maketitle

\abstract{We present a detailed analytic study on the three-dimensional sphere of the most popular candidates for $C$-functions, both for Dirac and scalar free massive fields. We discuss to which extent the effective action, the R\'enyi entanglement entropy and the renormalized entanglement entropy fulfill the conditions expected from $C$-functions. In view of the absence of a good candidate in the case of the scalar field, we introduce a new candidate, which we call the modified effective action, and analyze its pros and cons.
}

\vfill

\pagebreak
\section{Introduction}

One of the main difficulties Quantum Field Theory deals with is the large number of degrees of freedom confined to regions of very high energy scales. At each point of space there is a quantum field that fluctuates independently. As a consequence, one encounters ultraviolet (UV) divergences. There are ways to deal with such divergencies, and extract physically meaningful finite results which, in general, depend on the scale, making evident that physics may be different at different length scales that can vary from the UV to the infrared (IR); so, one still needs to interpret the scaling process. The authors of reference \cite{wilson} made use of the functional integral to approach this problem by integrating out degrees of freedom corresponding to high momenta or very small lengths. The result of such integration is an effective theory that involves quantum fields with momenta smaller than a fraction of the chosen cutoff. As a consequence, all the parameters of the theory, such as masses or coupling constants, become scale dependent and obey differential equations that are called the renormalization group equations. The foundations of the renormalization group can be traced back to 1953 with the work of Stueckelberg and Petermann \cite{stueckelberg_petermann}.

\begin{figure}[b]
\begin{center}
\includegraphics[scale=.5]{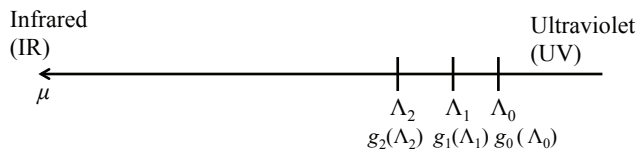}
\caption{\small Scales along the Renormalization Group flow}
\label{figureRGFlow}
\end{center}
\end{figure}

In order to be more specific, let us detail the process a little further. Following Kogut and Wilson, we can describe physics at different scales by making use of effective theories in a very intuitive way. We set up a UV cutoff $\Lambda_0$ above which the quantum fluctuations of the fields are ignored, and write down the Lagrangian of the theory, which now becomes an effective Lagrangian. This should contain a term corresponding to the free case plus other terms specific to the energy scale we are considering; these terms include the parameters that define the theory. Let us now lower the energy scale so that the new cutoff energy becomes $\Lambda_1<\Lambda_0$. In order to do so, we integrate out those fields whose momenta fall on the range of energies $[\Lambda_1,\Lambda_0]$. This gives rise to a new effective Lagrangian where the parameters differ from the previous ones, since the scale has changed, but we should still be able to recover the limit of the free theory. By iterating this process we obtain a series of effective theories with different parameters, say $g_0, g_1, g_2,....$, that are scale dependent (see Figure \ref{figureRGFlow}). This is called Renormalization Group.
The method involves a flow of the parameters $g_i$ running from the UV to the IR.

\begin{figure}[h]
\begin{center}
\includegraphics[scale=.5]{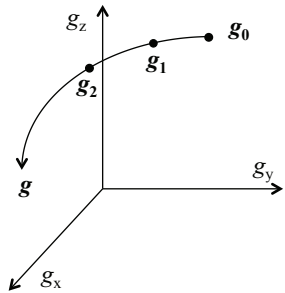}
\caption{\small Flow of the RG.}
\label{figureRG_g_Flow}
\end{center}
\end{figure}

As a consequence, theories along the flow are defined by effective coupling constants, resulting from renormalizing the theory at each scale, and they form a trajectory of the RG flow (see Figure \ref{figureRG_g_Flow}) that can be parametrized by 
$t=-$log$\,\mu$, that grows towards the IR, where $\mu$ is the energy scale. The renormalization group is generated by the vector field
  \begin{equation}
    \frac{d g^i}{dt}\equiv\mu\frac{d g^i(\mu)}{d\mu}=\beta^i(g(\mu)),\nonumber
  \end{equation}
where $\beta$ depends explicitly on the $g$'s but not on $\mu$.
The flow of the RG is, therefore, a one-parameter motion \cite{wilson} in the space of renormalized coupling constants, where beta functions appear as velocities,
  \begin{equation}
    \frac{\partial}{\partial t} = -\beta^i(g)\frac{\partial}{\partial g_i}.\nonumber
  \end{equation}
  Intuitively, one can think of the RG flow as a one-way process along which degrees of freedom are getting lost as we integrate out the configurations of the fields with higher and higher momenta. In particular, big masses get decoupled when the distance increases as we approach the IR.\\
RG flows have the so-called \emph{fixed points}, where the $\beta$ function becomes zero,
  \begin{equation}
    \mu\frac{d g^i}{d\mu}=0. \nonumber
  \end{equation}
  At such points, $g^i=g^*$ does not depend on the scale and the theory is scale invariant. This, together with Lorentz invariance, leads to a conformal field theory (CFT) at the fixed points.

In two dimensions the well-known Zamolodchikov's $c$-theorem \cite{zam} states that there exists a function $c(g)$ that decreases monotonically towards the IR and, at the fixed points, equals the central charge $c$ of the corresponding conformal theory. For conformal theories in two dimensions, many physically meaningful quantities were shown to be proportional to $c$. Among them, one can mention the trace of the energy-momentum tensor of the theory on a curved space---which leads to the well-known trace anomaly---, the entropy density at finite temperature in flat space and the universal part of the von Neumann entanglement entropy of an interval of length L, as well as the universal part of the entropy of an infinite system in a thermally mixed state \cite{cardy}.  
Zamolodchikov's $c$-theorem turned out to be an important tool in the study of two-dimensional field theories. It has been applied to statistical-mechanical models, as well as to the study of the topology of the space of all CFTs. In view of this, extensions to higher dimensions started to be sought for. After many attempts at generalizing the $c$-theorem to even-dimensional theories \cite{cardy4, osborn, osborn-jack}, a proof for the four-dimensional case has recently emerged \cite{KS,K}.  In this case, there exists a function with a suitable behavior under the RG flow that coincides at the conformal fixed points with the coefficient $a$ of the conformal anomaly. In this dimension, there are examples where the anomaly coefficient $c$ does not, instead, have the desired behavior \cite{Cappelli:1990yc,Anselmi:1997am}.

It seems that in even dimensions, where we have the anomaly of the trace, it is possible to relate the $\textit{C}$-function that runs under the RG flow to some coefficient of this anomaly.
Things are less clear in odd dimensions, where not even a conformal anomaly does exist.

If we want to find a function that describes the flow towards the IR of the RG of a theory, we need to state what we should expect from such function. In general, there are different levels of requirements that a physical property should satisfy in order to be considered a $\textit{C}$-function \cite{Barnes2004}:

\begin{enumerate}
	\item On one side, one can think of a quantity $C$ that counts the degrees of freedom of the CFTs connected by a RG flow: $C\geq 0$ with $C_{IR}<C_{UV}$. \label{r1}
	\item Moreover, one can seek for a function that interpolates monotonically between the CFT values, thus giving a measure of the degrees of freedom for all the theories along the flow:
	\[\dot{C}(g(t)) = -\beta^{i}(g)\frac{\partial}{\partial {g}_i}C(g(t))\leq 0\,,\] with $\dot{C}= 0$ $iff$ the theory is conformal. \label{r2}
	\item Finally, one can ask for the strongest condition \[\dot{C}(g(t)) = -G_{ij}\beta^{i}(g)\beta^{j}(g)\,,\] with $G_{ij}$ a positive definite matrix, which implies the previous condition. \label{r3}
\end{enumerate}

In this paper, we investigate theories on the three-dimensional sphere and discuss to which extent some candidates for $\textit{C}$-functions fulfill the requirements listed above. Always in a $\zeta$-regularization framework, we give fully analytical results for them in the cases of massive free scalar and Dirac fields.
	
The paper is organized as follows: We start, in section \ref{sect2} with the effective action for massive free fields on $S^3$.
In the case of the Dirac field, we find that the effective action does satisfy the two first conditions above and is, thus, a good candidate for a $\textit{C}$-function. Regarding the scalar field, we obtain an IR divergence, which cannot be renormalized away without breaking unitarity or leading to a not always positive nor monotonic function, with a divergent derivative with respect to the coupling constant at the UV fixed point. In view of this, in section \ref{sect3} we study the R\'enyi entanglement entropy for the same theories, using the replica trick \cite{cardy}. In the scalar case we still find an IR divergence, proportional to the volume of the codimension two entanglement region, which is also difficult to renormalize. We do not find major improvement in the Dirac field case either. In section \ref{sect4} we discuss the so-called renormalized entanglement entropy, which we find in the case of the scalar field far from being a good $C$-function, since it also becomes negative in some ranges of the coupling constant. For the Dirac field, instead, it fulfills all the requirements listed above. Finally we present, in section \ref{sect5}, a new candidate for a $\textit{C}$-function that we call a modified effective action. With this function we are able to fix most of the problems found in previous candidates except that it is still unstable at the UV fixed point.

Additionally, new useful identities regarding series of zeta functions, which we use to perform our analytic calculations, are proved in the appendix \ref{ap2}.

\section{The effective action on $S^3$}\label{sect2}

This problem was studied in \cite{Jafferis:2011zi,kleF}, where the validity of the so-called $F$-theorem was con-jectured---the statement that the effective action $F$ of any theory on the three-sphere at its conformal fixed points satisfies condition \ref{r1}, i.e., $F_{UV}\geq F_{IR}$. The authors of these references showed that $F$ is UV-stable in the case of free scalar fields. This quantity was shown to coincide, at the same fixed point, with the universal part of the entanglement entropy across a circle \cite{Casini:2011kv}, which was proven to be monotonically decreasing along any flow for unitary, relativistic theories \cite{casini}. As noticed in \cite{kleF}, when one analyzes the mass flow for a free scalar or fermion theory starting from $F_{UV}$ and using $F$ on the sphere as the $C$-function over the whole mass range, there is an IR divergence (also discussed in \cite{dowkermasssphere}), which cannot be subtracted without breaking the unitarity of the theory or getting a not always positive function, with a divergent derivative with respect to the ``coupling constant'' at the UV fixed point.

In this section we perform an analytic computation of $F$ in the whole range of variation of the dimensionless coupling, both for scalar and Dirac fields. In the last case, we find that the effective action does satisfy conditions \ref{r1} and \ref{r2}, which makes it a good candidate for a $\textit{C}$-function. It fails to satisfy, however, the more stringent stability condition \ref{r3} at the UV fixed point. Regarding the scalar theory, we obtain the expected IR-divergent behavior of $F$. As commented in \cite{kleF,dowkermasssphere}, we find that subtracting the full divergent contribution to $F$ would break the unitarity of the theory near the UV. We also show that a subtraction of only the leading divergencies gives neither a monotonic nor a positive function, and leads to a divergent derivative with respect to the coupling constant at the UV fixed point.

\subsection{Conformally coupled massive scalar field}

The eigenvalues corresponding to the conformally coupled massive Laplacian on $S^3$ are given by
\beq
\lambda_n = \frac{n^2 - 1/4}{a^2} + m^2\,,\qquad n=1,2,\ldots,\infty
\eeq
with degeneracies
\beq
d_n  =n^2\,.
\eeq

Thus, the corresponding zeta function is given by
\beq
\zeta^{S^3}(s)=
(\mu a)^{2s} \sum_{n=1}^{\infty} n^2 \left(n^2 -\frac{1}{4} + (ma)^2\right)^{-s} \,,
\eeq
where $\mu$ was introduced to render the zeta function dimensionless. However, it is well-known that $\zeta^{S^3}(s=0)$ vanishes. So, from now on, we will disregard the pre-factor $(\mu a)^{2s}$.

We start by considering the behavior of the effective action near the UV fixed point. Let's call ${\rho}^2 =\frac{1}{4} - (ma)^2 $. For $ |\rho |< 1$, the binomial expansion leads to
\beq
\zeta^{S^3}(s)&=&
\sum_{n=1}^{\infty} n^{2-2s} \sum_{j=0}^{\infty} \frac{\Gamma(-s+1)}{j!\,\Gamma(-s+1-j)} (-1)^j {\rho}^{2j} n^{-2j}\nn \\
&=& \sum_{j=0}^{\infty} \frac{\Gamma(-s+1)}{j!\,\Gamma(-s+1-j)} (-1)^j {\rho}^{2j} \zeta_R (2s+2j-2)
\nn \\
&=& \zeta_R (2s-2) + \sum_{j=1}^{\infty} \frac{\Gamma(s+j)}{j!\,\Gamma(s)} {\rho}^{2j} \zeta_R (2s+2j-2)\,.
\eeq

The effective action for a real scalar field in this range of masses is given by
\beq
S_{\mathrm{eff}}^{S^3} = \left. -\frac12 \frac{d}{ds} \right\vert_{s=0} \zeta^{S^3}(s) = - \zeta^{\prime}_R (-2) - \frac12 \sum_{j=0}^{\infty} \frac{{\rho}^{2j +2}}{j+1} \zeta_R (2j)\,.
\eeq

Using a result from reference \cite{sri}, the above can be evaluated to give
\beq
S_{\mathrm{eff}}^{S^3} &=&-\frac12 \left\{ {\rho}^{2} \,\zeta^{\prime}_H (0, 1-\rho) +
2 {\rho} \,\zeta^{\prime}_H (-1, 1-\rho)+\zeta^{\prime}_H (-2, 1-\rho)+(\rho\rightarrow -\rho)\right\}\,,
\label{seffs3}\eeq
where we have introduced the notation $(\rho\rightarrow -\rho)$, meaning that the same terms changing $\rho$ by $-\rho$ have to be added.

In particular, at the UV fixed point, the effective action reduces to the positive value $\frac18 [\log(2) - \frac{3 }{2 {\pi}^2}\zeta(3)]$, which agrees with the result in \cite{kleF}. It is well-known that the derivative with respect to $(ma)^2$ vanishes at this same conformal fixed point. In summary, the properties of the effective action at the UV fixed point are compatible with those of a  $\textit{C}$-function at the same fixed point.

However, near the IR we do not find such good behavior. Indeed, an alternative expression of the effective action, more suitable for studying the IR region, can be obtained by inverting the full heat kernel; after doing so we get
\beq  \zeta^{S^3}(s) &=& \sum_{n=1}^{\infty} n^2 (n^2 + (ma)^2 -\frac{1}{4})^{-s}\nn \\
&=& \sum_{n=0}^{\infty} (n^2 + {\delta}^2 )^{-s+1} - {\delta}^2 \sum_{n=1}^{\infty} (n^2 + {\delta}^2 )^{-s}\nn \\
&=& \frac12 \left\{ \sum_{n=- \infty}^{\infty} (n^2 + {\delta}^2 )^{-s+1} - {\delta}^2 \sum_{n=-\infty}^{\infty} (n^2 + {\delta}^2 )^{-s}\right\}
\,, \nn\eeq
where we have defined ${\delta}^2= (ma)^2 - \frac14$.

After performing a Mellin transform and using the inversion formula for the Jacobi Theta function, we obtain
\beq
\zeta^{S^3}(s) &=& \frac{{\pi}^{\frac12} (s-1)}{2 \Gamma(s)} \left[\Gamma(s-\tfrac32) ({\delta}^2 )^{\frac32-s} +
4 \sum_{n=1}^{\infty} \left(\frac{n \pi}{\delta}\right)^{s-\frac32} K_{s-\frac32}(2 \delta n \pi)\right]\nn \\
& &-
\frac{{\pi}^{\frac12} {\delta}^2}{2 \Gamma(s)} \left[\Gamma(s-\tfrac12) ({\delta}^2 )^{\frac12-s} +
4 \sum_{n=1}^{\infty} \left(\frac{n \pi}{\delta}\right)^{s-\frac12} K_{s-\frac12}(2 \delta n \pi)\right]\,.
\eeq
The effective action then reads
\beq
S_{\mathrm{eff}}^{S^3} = - \frac12 &&\left\{-\frac{{\pi}^{\frac12}}{2} \left[\Gamma(-\tfrac32) ({\delta}^2 )^{\frac32} +
4 \sum_{n=1}^{\infty} \left(\frac{n \pi}{\delta}\right)^{-\frac32} K_{\frac32}(2 \delta n \pi)\right]\right.\nn \\
 &&-\left.\frac{{\pi}^{\frac12} {\delta}^2}{2} \left[\Gamma(-\tfrac12) ({\delta}^2 )^{\frac12} +
4 \sum_{n=1}^{\infty} \left(\frac{n \pi}{\delta}\right)^{-\frac12} K_{\frac12}(2 \delta n \pi)\right]\right\}\,.
\label{Seff-S3-IR}\eeq

It is worth to point out that this result coincides with the analytic extension of the result in equation \eqref{seffs3} for $\rho=i\delta$.

\begin{figure}[h]
	\centering
	\begin{minipage}{.4\textwidth}
		\centering
		\includegraphics[height=42mm]{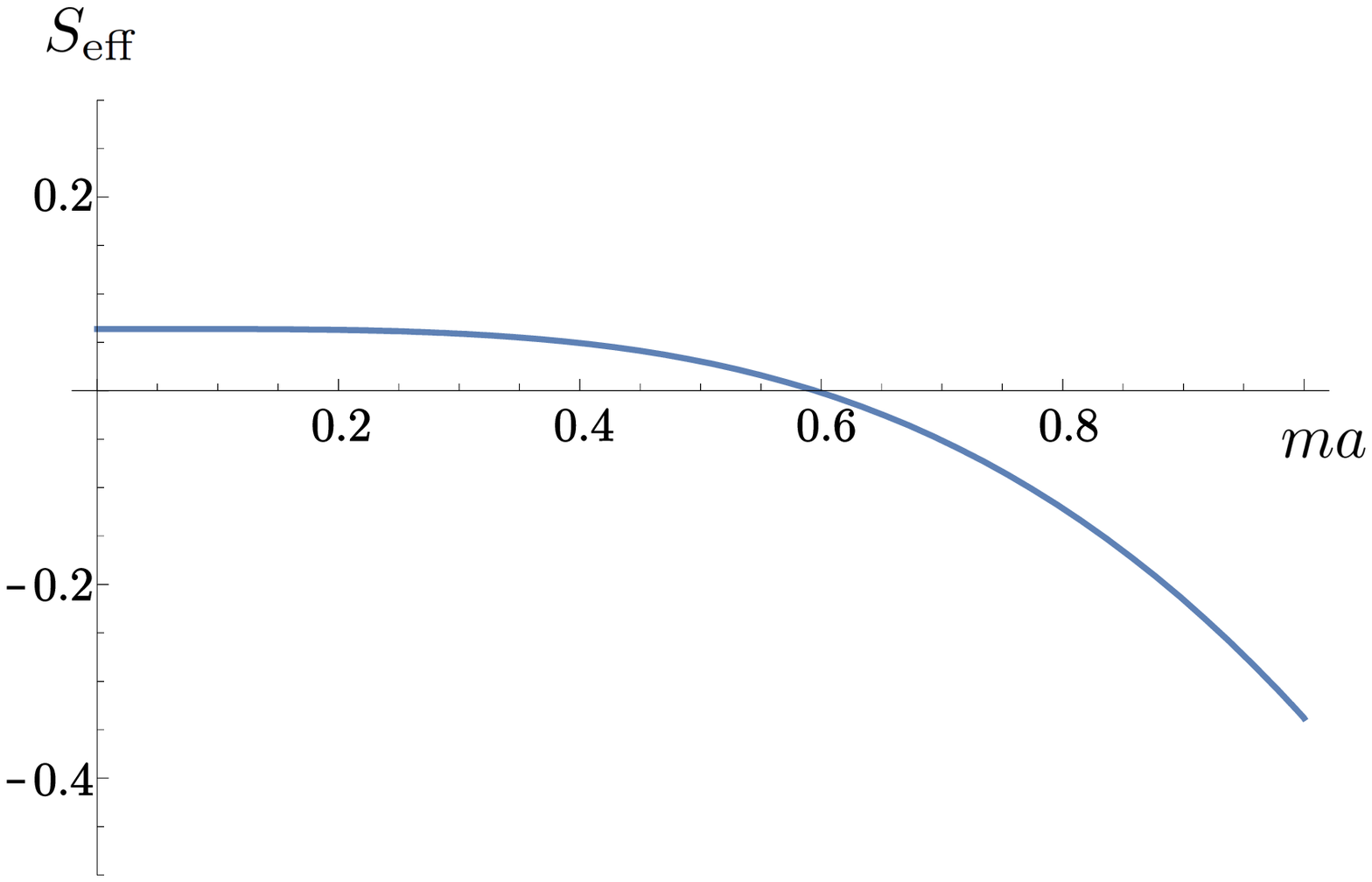}
		\caption{\small Effective action for a real scalar field on $S^3$ as a function of $ma$}
		\label{figure0}
	\end{minipage}
	\hspace{0.02\textwidth}
	\begin{minipage}{.4\textwidth}
		\centering
		\includegraphics[height=42mm]{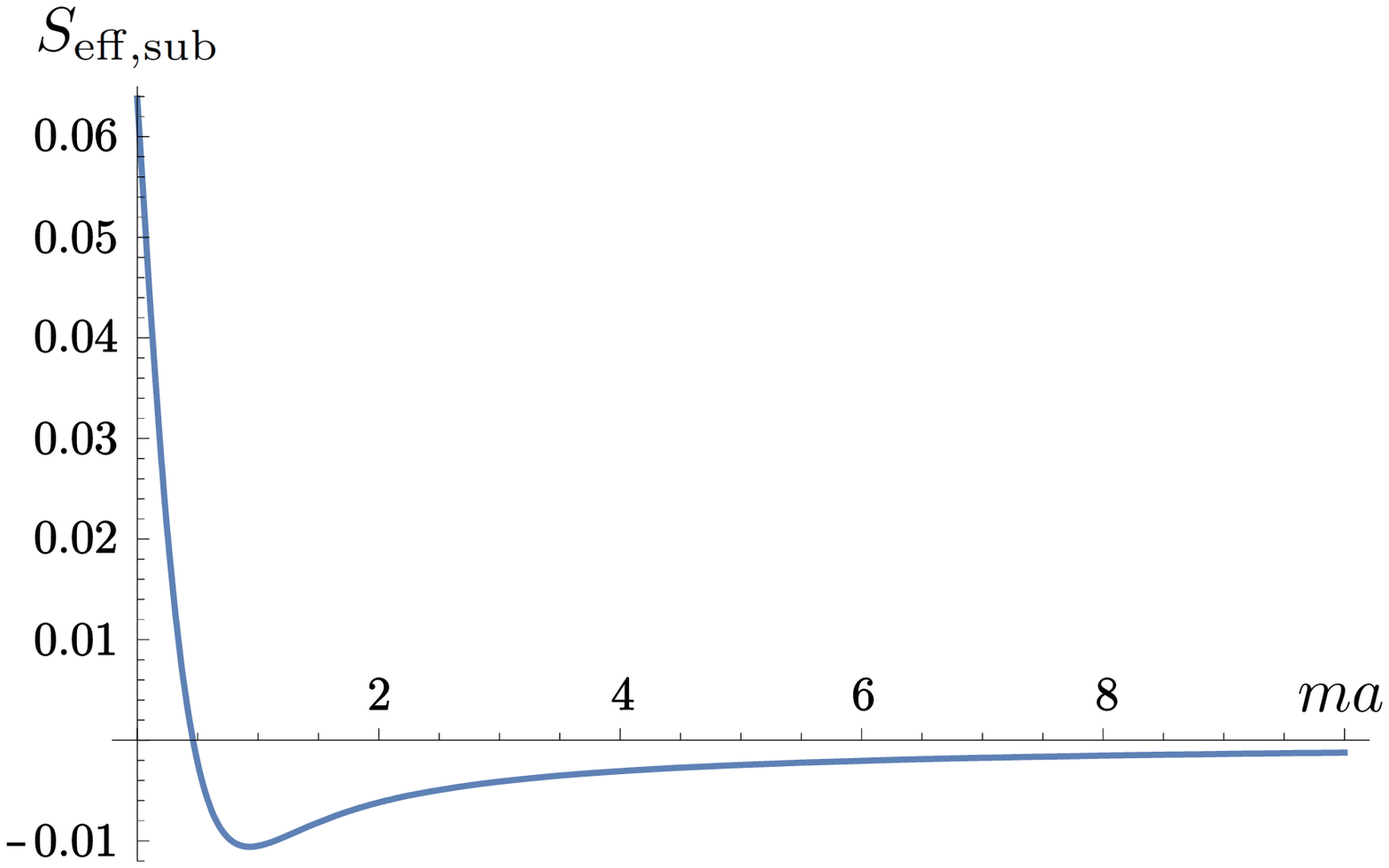}
		\caption{\small Effective action for a real scalar field on $S^3$ as a function of $ma$, where the leading IR divergence has been subtracted}
		\label{figjj}
	\end{minipage}
\end{figure}

In the IR limit $ma \rightarrow \infty$, the effective action in \eqref{Seff-S3-IR} asymptotes at
\beq
-\frac{\pi}{6} ({\delta}^2)^{\frac32}\,,\label{seffS3}
\eeq
which is divergent. Such behavior shows clearly in Figure \ref{figure0}.

As it is, the whole divergence cannot be subtracted through an adequate renormalization procedure, since it would lead to an imaginary contribution at the UV fixed point, therefore breaking unitarity. One could, instead, subtract only the dominant divergence ($-\frac{\pi}{6}[(ma)^3-\frac38 (ma)]$). As pointed in \cite{komleshouches}, this would amount to adding a cosmological constant counterterm plus an Einstein-Hilbert counterterm. However, doing so would lead to a not always positive nor monotonic function (see Figure \ref{figjj}), with a divergent derivative with respect to the dimensionless coupling $(ma)^2$ at the UV fixed point. So, it is clear that, although the effective action on the sphere fulfills all the required properties at the UV fixed point, the IR divergence rules it out as  candidate for a $\textit{C}$-function. In the next subsection, we show that the situation is entirely different in the case of the Dirac field.

\subsection{Massive Dirac field}

In order to study the behavior of the effective action for the Dirac field on the three-sphere under the mass RG flow, we start from the eigenvalues of the massive Dirac operator, which are
\beq
\lambda_n^{\pm}= \pm i\frac{(n+\frac{3}{2})}{a} + m\,, \qquad n= 0,1,\ldots , \infty
\eeq
with degeneracies
\beq
d_n= (n+1)(n+2)\,,
\eeq
and build the corresponding zeta function,
\beq
\zeta^{S^3}(s) &=& (\mu a)^{s} \left\{e^{-i\frac{\pi}{2}s} \sum_{n=0}^{\infty} n(n+1) \left(n+\frac12 - i ma\right)^{-s} + e^{i\frac{\pi}{2}s} \sum_{n=0}^{\infty} n(n+1) \left(n+\frac12 + i ma\right)^{-s}\right\} \nn \\
&=&(\mu a)^{s}  \Bigg\{e^{i\frac{\pi}{2}s}\Bigg[\zeta_H\left(s-2, \frac12 + i ma\right) - 2 i ma\, \zeta_H\left(s-1, \frac12 + i ma\right)
    \nn\\
&&\qquad\qquad\qquad- \left(\frac14 + (ma)^2\right)\zeta_H\left(s, \frac12 + i ma\right)\Bigg]
  + e^{-i\frac{\pi}{2}s}\left[(ma\rightarrow -ma)\right]\Bigg\}\,.\label{zeta-Dirac}
\eeq

It is easy to check that, as in the scalar case, this $\zeta$ function vanishes at $s=0$. Therefore, the effective action is given by the real expression
\beq
S_{\mathrm{eff}}^{S^3} &=& \left.\frac{d}{ds} \right\vert_{s=0} \zeta^{S^3}(s) \nn \\
&=&- \frac{\pi}{12} ma \left[3 + 4 (ma)^2\right] + \Bigg{\{}\zeta^{\prime}_H\left(-2, \frac12 + i ma\right)- 2 i ma\,\zeta^{\prime}_H\left(-1, \frac12 + i ma\right)\nn\\
&&- \left[\frac14 + (ma)^2\right]\zeta^{\prime}_H\left(0, \frac12 + i ma\right) +(ma\rightarrow -ma)\Bigg{\}}\,.
\label{seffdirac}
\eeq

Notice that the first term comes from the spectral asymmetry when one selects, as we do here, the decoupling convention for the phase of the determinant \cite{jackiw}. Thus, it vanishes at $ma=0$. Such contribution was initially ignored, for instance, in \cite{kleF}. The value of the effective action at the UV fixed point is $ \frac{\log(2)}{4}+\frac{3}{8 \pi^2} \zeta_R(3)$, which is positive. However, its derivative with respect to $ma$, which is the dimensionless coupling constant for this theory, fails to vanish at the same point, where it is given by $-\frac{\pi}{4}$ due, precisely, to the contribution of the asymmetry. Indeed, it is easy to see that
\beq
\frac{d}{d(ma)}S_{\mathrm{eff}}^{S^3}=\pi\,\left[\frac14+(ma)^2\right]\left[\tanh(\pi\,ma)-1\right]\,.\label{derdirac}
\eeq

So, as a function of $ma$, the effective action is decreasing. Moreover, one can check, by using the asymptotic expansions for the derivatives of the Hurwitz zeta functions in reference \cite{elizalde}, that it vanishes in the IR limit, which would not be the case if the contribution from the spectral asymmetry had been ignored or the opposite phase convention had been chosen. In that case, as noted in \cite{kleF}, some subtraction would be needed.

\begin{figure}[h]
	\centering
\includegraphics[height=42mm]{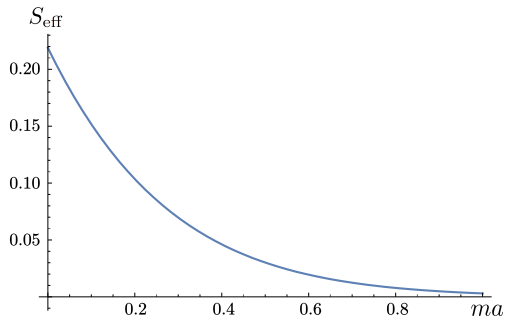}
\caption{\small Effective action for a Dirac field on $S^3$ as a function of $ma$}
\label{figure1}
\end{figure}

The behavior of the effective action for a Dirac field on $S^3$ is shown in Figure \ref{figure1}.

\section{R\'enyi entanglement entropy}\label{sect3}

Since the effective action on $S^3$ coincides at the UV fixed point with minus the R\'enyi entanglement entropy it is natural to check whether this last function behaves as expected from a $\textit{C}$-function. In this section we study, using the expression that follows from the replica trick, its behavior under the flow of the renormalization group, again for free massive real scalars and fermions.

The R\'enyi entanglement entropy for multiple ($q$) coverings of $S^3$ is given by
\beq
S=\lim _{q\rightarrow 1} S_q=\lim _{q\rightarrow 1} \left( \frac{q S_{\mathrm{eff}}^{S^3} - S_{\mathrm{eff}}^{qS^3}}{1-q} \right)\,,
\label{Renyi}
\eeq
where $ S_q$ is the R\'enyi entropy of order $q$ and $S_{\mathrm{eff}}^{S^3}$ is the effective action for a conformally coupled field on $S^3$, while $S_{\mathrm{eff}}^{qS^3}$ is the effective action for the field on the $q$-covering of the same manifold.

\subsection{Massive scalar field}

The eigenvalues and degeneracies of the massless conformal Laplacian in three dimensions for multiple ($q$) coverings of $S^3$ were given in reference \cite{kleRenyi}. In the massive case, they are
\beq
\lambda_n = \frac{n^2 - 1/4}{a^2} + m^2\,,
\eeq
with degeneracies
 \beq
 d_n  =\begin{cases} n^2,&  \quad \mbox{for} \quad n=1,2,..., \infty \\
      k(k+1),& \quad \mbox{for} \quad  n= k+\frac{p}{q} \quad k=1,2,..., \infty\,, \quad p=1,2,...,q-1\end{cases}\,.
\eeq

Thus, the corresponding zeta function is given by
\beq
\zeta^{qS^3}(s)&=& (a \mu)^{2s}\Bigg\{
\sum_{n=1}^{\infty} n^2 \left(n^2 -\frac{1}{4} + (ma)^2\right)^{-s}\nn \\ && \qquad\qquad\qquad+ \sum_{k=1}^{\infty}\sum_{p=1}^{q-1} k(k+1) \left(\left(k + \frac{p}{q}\right)^2 -\frac{1}{4} + (ma)^2\right)^{-s}\Bigg\} \nn \\
&=& \zeta_1 (s) + \zeta_2 (s)\,.\label{zetaq}
\eeq

We begin by studying the behavior of the entanglement entropy near the UV through a detailed evaluation of the relevant effective action for small values of $ma$. By doing so, we determine, by analytical continuation to continuous values of $q$, the relation between the entanglement entropy and the effective action on the sphere. From such relation, we verify that such entropy does, indeed, coincide with minus the effective action on $S^3$ in the UV limit, as stated by the $F$-theorem and proved, through a different approach, in \cite{dowkerentang}. Moreover, we also show that the UV fixed point is not stable, i.e., $\frac{d}{d(ma)^2}\left.S\right\vert_{ma=0}\neq 0$, as in \cite{dowkerrenyi}. Finally, by making use of an adequate analytic continuation to the whole range of $ma$, we analyze the behavior of the entanglement entropy in the IR limit and reproduce its divergent behavior at the IR fixed point.

The first term in equation \eqref{zetaq} ($\zeta_1$) gives a contribution to the effective action which coincides with the result in \eqref{seffs3}
\beq
S_{\mathrm{eff},1}&=&-\frac12 \left\{ {\rho}^{2} \,\zeta^{\prime}_H (0, 1-\rho)  +
2 {\rho} \,\zeta^{\prime}_H (-1, 1-\rho) + \zeta^{\prime}_H (-2, 1-\rho) + (\rho\rightarrow -\rho)\right\}\,.
\label{seff1}\eeq

Then, we only need to evaluate the $\zeta_2$ term, which gives the contribution coming from noninteger eigenvalues.

As in the previous section, we first consider the behavior near the UV fixed point, $(ma)^2< \frac{1}{4}$. In the zero mass case \cite{dowkerentang}, there is no multiplicative anomaly in this problem and it can easily be shown that this is also true in the range of $\rho$ we are considering. This allows us to obtain the effective action from an equivalent $\zeta_2$ that we call $\tilde\zeta_2$,
\beq
\tilde{\zeta}_2= \sum_{k=1}^{\infty}\sum_{p=1}^{q-1} k(k+1) \left(\left(k + \frac{p}{q}\right) - \rho \right)^{-s} +
\sum_{k=1}^{\infty}\sum_{p=1}^{q-1} k(k+1) \left(\left(k + \frac{p}{q}\right) + \rho \right)^{-s}\,.
\eeq

After using the binomial expansion, it reads
\beq
\tilde{\zeta}_2(s) &=& 2 \sum_{k=1}^{\infty}\sum_{p=1}^{q-1}\sum_{j=0}^{\infty} k(k+1) \left(k + \frac{p}{q}\right)^{-s-2j} \frac{\Gamma(-s+1)}{(2j)!\,\Gamma(-s+1-2j)} {\rho}^{2j}\nn\\
&=&  2 \sum_{p=1}^{q-1}\sum_{j=0}^{\infty}  \frac{\Gamma(-s+1)}{(2j)!\,\Gamma(-s+1-2j)} {\rho}^{2j} \left[ \zeta_H (s+2j-2, 1+\frac{p}{q})\right.\nn \\
&&+ \left.  \left( 1-\frac{2p}{q}\right)
\zeta_H (s+2j-1, 1+\frac{p}{q}) - \frac{p}{q}\left( 1-\frac{p}{q}\right)\zeta_H (s+2j, 1+\frac{p}{q}) \right]\,.
\eeq

Its contribution to the effective action is thus given by
\beq
S_{\mathrm{eff},2} &=&-\frac12 \left.\frac{d}{ds} \right\vert_{s=0} \tilde{\zeta}_2(s) \nn \\
&=& - \sum_{p=1}^{q-1} \left[ \zeta^{\prime}_H (-2, 1+\frac{p}{q}) + \left( 1-\frac{2p}{q}\right)
\zeta^{\prime}_H (-1, 1+\frac{p}{q}) - \frac{p}{q}\left( 1-\frac{p}{q}\right)\zeta^{\prime}_H (0, 1+\frac{p}{q})\right. \nn \\
&&+ \frac{{\rho}^2}{2} \left( 1-\frac{2p}{q}\right) \left(1-\psi(1 + \frac{p}{q})\right) + \sum_{j=1}^{\infty} \frac{{\rho}^{2j}}{2j} \zeta_H (2j-2, 1+\frac{p}{q}) \\
&&-\left. \frac{p}{q}\left( 1-\frac{p}{q}\right) \sum_{j=1}^{\infty} \frac{{\rho}^{2j}}{2j} \zeta_H (2j, 1+\frac{p}{q}) +
\left( 1-\frac{2p}{q}\right) \sum_{j=1}^{\infty} \frac{{\rho}^{2j+2}}{2j+2} \zeta_H (2j+1, 1+\frac{p}{q}) \right]\,.\nn
\eeq

Now, using a result from \cite{sri}, we get
\beq
S_{\mathrm{eff},2} &=& - \sum_{p=1}^{q-1} \Bigg\{ \frac12\, \zeta^{\prime}_H (-2, 1+\frac{p}{q}- \rho) + \left[ \frac12\left( 1-\frac{2p}{q}\right) +\rho \right] \zeta^{\prime}_H (-1, 1+\frac{p}{q}- \rho)
\nn \\
&+& \left[\frac{{\rho}^2}{2} -\frac12 \frac{p}{q}\left( 1-\frac{p}{q}\right)+\frac{{\rho}}{2}\left( 1-\frac{2p}{q}\right)\right] \zeta^{\prime}_H (0, 1+\frac{p}{q}-\rho)]+(\rho\rightarrow -\rho)\Bigg\}\,.
\label{seff2}
\eeq

One can check that, in the UV limit ($\rho=\frac12$), the sum of $S_{\mathrm{eff},1}$ and $S_{\mathrm{eff},2}$ reduces to the result in equation (3.12) of reference \cite{kleRenyi}.

Now, from the definition of the entanglement entropy (equation \eqref{Renyi}), and using \eqref{zetaq} along with the relation between the effective action and its corresponding zeta function, we find
\beq
S= - S_{\mathrm{eff}}^{S^3} -\lim_{q\rightarrow 1} \frac{S_{\mathrm{eff},2}}{1-q}\,. \label{srenyiesc}
\eeq

As already said, we obtain $S$ by performing an analytic extension of the last term in this equation to continuous values of $q$. In order to do this, we  make use of several results listed in Appendix \ref{ap2}. Equation \eqref{seff2} leads us to
\beq
\lim_{q\rightarrow 1} \frac{- S_{\mathrm{eff},2}}{1-q} &=& \lim_{q\rightarrow 1} \frac{1}{1-q} \sum_{p=1}^{q-1} \left\{ \frac12 \left[\zeta^{\prime}_H (-2, 1+\frac{p}{q}- \rho) + \zeta^{\prime}_H (-2, 1+\frac{p}{q}+ \rho)\right] \right. \nn \\
&& + \left[ \frac12\left( 1-\frac{2p}{q}\right) +\rho \right] \zeta^{\prime}_H (-1, 1+\frac{p}{q}- \rho) + \left[\frac12\left( 1-\frac{2p}{q}\right)-\rho \right] \zeta^{\prime}_H (-1, 1+\frac{p}{q}+ \rho)
\nn \\
&& + \left[\frac{{\rho}^2}{2} -\frac12 \frac{p}{q}\left( 1-\frac{p}{q}\right)+\frac{{\rho}}{2}\left( 1-\frac{2p}{q}\right)\right] \zeta^{\prime}_H (0, 1+\frac{p}{q}-\rho)\nn \\
&& + \left.\left[\frac{{\rho}^2}{2} -\frac12 \frac{p}{q}\left( 1-\frac{p}{q}\right)-\frac{{\rho}}{2}\left( 1-\frac{2p}{q}\right)\right] \zeta^{\prime}_H (0, 1+\frac{p}{q}+\rho)
\right\}\nn\\
&=& \lim_{q\rightarrow 1} \frac{1}{1-q}\left(A(q,\rho)+B(q,\rho)+C(q,\rho)\right)\,.
\eeq

By using the multiplication formula reproduced in \eqref{a21}, the sum $A(q,\rho)$ can be rewritten as
\beq A(q,\rho)=\left.\frac{d}{ds}\right\vert_{s=-2} \frac{1}{2}\left[ q^s \zeta_H (s, q(1-\rho)) + q^s \zeta_H (s, q(1+\rho))- \zeta_H (s,1-\rho )- \zeta_H (s,1+\rho )\right]\nn\,. \eeq
Now, it is easy to see that the expression between square brackets in the equation above vanishes for all $s$ and for all $\rho$ at $q=1$. Thus
\beq
\lim_{q\rightarrow 1} \frac{A(q,\rho)}{1-q} &=&  -\frac12 \left.\frac{d}{ds}\right\vert_{s=-2} \left.\frac{d}{dq}\right\vert_{q=1}\left[ q^s \zeta_H (s, q(1-\rho)) + q^s \zeta_H (s, q(1+\rho))\right.\nn\\
&& -\left. \zeta_H (s,1-\rho )- \zeta_H (s,1+\rho )\right] \nn\\
& = &  -\frac12 \left[\zeta_H (-2, 1-\rho)-2\zeta_H^{\prime} (-2, 1-\rho)+\zeta_H (-2, 1+\rho)-2\zeta_H^{\prime} (-2, 1+\rho)\right.\nn\\
&& - (1-\rho)\zeta_H (-1, 1-\rho)+2(1-\rho)\zeta_H^{\prime} (-1, 1-\rho)\nn\\
&& - \left. (1+\rho)\zeta_H (-1, 1+\rho)+2(1+\rho)\zeta_H^{\prime} (-1, 1+\rho)\right]\,, \label{aesc}
\eeq
where we have used \eqref{a23} to obtain the last equality.

We now extend $B(q,\rho)$; we start by writing the finite sum as the difference of two infinite series,
\beq
B(q,\rho)&=&\left.\frac{d}{ds}\right\vert_{s=-1} \sum_{p=1}^{q-1}\left[\left(\frac12 -\frac{p}{q}+\rho \right)\zeta_H (s, 1+\frac{p}{q}-\rho) +
\left(\frac12 -\frac{p}{q}-\rho \right)\zeta_H (s, 1+\frac{p}{q}+\rho)\right]\nn\\
&=& \left.\frac{d}{ds}\right\vert_{s=-1}\left\{ \sum_{p=1}^{\infty}\left(\frac12 -\frac{p}{q}+\rho \right)\zeta_H (s, 1+\frac{p}{q}-\rho)
\right.\nn\\
&& - \left.\sum_{p=q}^{\infty}\left(\frac12 -\frac{p}{q}+\rho \right)\zeta_H (s, 1+\frac{p}{q}-\rho) + (\rho \rightarrow -\rho) \right\}\,,\nn
\eeq
and change indices to obtain
\beq
B(q,\rho)=  \left.\frac{d}{ds}\right\vert_{s=-1} && \left\{ \sum_{p=1}^{\infty}\left(\frac12 -\frac{p}{q}+\rho \right)\zeta_H (s, 1+\frac{p}{q}-\rho)
\right.\nn\\
&& - \left.\sum_{p=1}^{\infty}\left(\frac12 -\frac{p}{q}+\frac{1-q}{q}+\rho \right)\zeta_H (s, 1+\frac{p}{q}-\frac{1-q}{q}-\rho) + (\rho \rightarrow -\rho) \right\}\,.\nn
\eeq
Notice that the expression between curly brackets in the previous equation vanishes at $q=1$ for all $s$ and for all $\rho$. So
\beq
\lim_{q\rightarrow 1} \frac{B(q,\rho)}{1-q} &=& -\left.\frac{d}{ds}\right\vert_{s=-1} \left.\frac{d}{dq}\right\vert_{q=1}\left\{\sum_{p=1}^{\infty}
\left(\frac12 -\frac{p}{q}+\rho \right)\zeta_H (s, 1+\frac{p}{q}-\rho)\right.\nn\\
&& - \left.\sum_{p=1}^{\infty}\left(\frac12 -\frac{p}{q}+\frac{1-q}{q}+\rho \right)\zeta_H (s, 1+\frac{p}{q}-\frac{1-q}{q}-\rho) + (\rho \rightarrow -\rho) \right\}\nn\\
&=& -\left.\frac{d}{ds}\right\vert_{s=-1}\left\{\sum_{p=0}^{\infty}\zeta_H (s, 2+p-\rho)\right.\nn\\
&&\qquad\qquad\qquad+ \left.s \sum_{p=0}^{\infty}\left(\rho-\frac12 -p\right)\zeta_H (s, 2+p-\rho)+ (\rho \rightarrow -\rho)\right\}\,,\nn
\eeq
where we have used \eqref{a22} to obtain the last expression. Now, making use of \eqref{a25} and \eqref{a26}, we finally get
\beq
\lim_{q\rightarrow 1} \frac{B(q,\rho)}{1-q} &=&-\Bigg\{ \frac32\,\zeta_H^{\prime}\, (-2, 1-\rho)- (2-\rho)\,\zeta_H^{\prime} (-1, 1-\rho)\nn\\
&& +\frac12\,(1-\rho^2)\,\zeta_H^{\prime} (0, 1-\rho)-\frac12\,\zeta_H(-2, 1-\rho)\nn\\
&& +\zeta_H(-1, 1-\rho)-\frac12\,(1-\rho^2)\,\zeta_H (0, 1-\rho)+(\rho \rightarrow -\rho)\Bigg\}\,.
\label{besc}
\eeq

In order to extend $C(q,\rho)$, we rewrite it as
\beq
C(q,\rho)&=&\left.\frac12 \frac{d}{ds}\right\vert_{s=0} \left\{\sum_{p=1}^{q-1}\left[\left(\frac{p}{q}-\frac12 -\rho \right)^2-\frac14 \right]\zeta_H\left(s, 1+\frac{p}{q}-\rho\right) +
(\rho \rightarrow -\rho)\right\}\,.\nn
\eeq
As done earlier, we write this expression as a difference of two series,
\beq
C(q,\rho)&=&\left.\frac12 \frac{d}{ds}\right\vert_{s=0} \left\{\sum_{p=1}^{\infty}\left[\left(\frac{p}{q}-\frac12 -\rho \right)^2-\frac14 \right]\zeta_H\left(s, 1+\frac{p}{q}-\rho\right) \right.\nn\\
&& - \left.\sum_{p=1}^{\infty}\left[\left(\frac{p}{q}+\frac{q-1}{q}-\frac12 -\rho \right)^2-\frac14 \right]\zeta_H\left(s, 1+\frac{p}{q}+\frac{q-1}{q}-\rho\right)+
(\rho \rightarrow -\rho)\right\}\,.\nn
\eeq
Again, the expression between curly brackets vanishes at $q=1$ for all $s$ and for all $\rho$. Following close steps to the ones that lead us to \eqref{aesc} and \eqref{besc}, and using \eqref{a23} together with \eqref{a25}, \eqref{a26} and \eqref{a27}, we obtain
\beq
\lim_{q\rightarrow 1} \frac{C(q,\rho)}{1-q} &=&-\frac12\Bigg\{-\zeta_H^{\prime}(-2,1-\rho)+2\,\zeta_H^{\prime}(-1,1-\rho)-(1-\rho^2)\,\zeta_H^{\prime}(0,1-\rho)\nn\\
&& +\frac13\,\zeta_H(-2,1-\rho)-\zeta_H(-1,1-\rho)+\frac23\,\zeta_H(0,1-\rho)\nn\\
&& +\frac13\,\rho\,(1-\rho^2)\,\Psi(1-\rho)+ (\rho \rightarrow -\rho)\Bigg\}\,.
\label{cesc}
\eeq

Now, putting \eqref{aesc}, \eqref{besc} and \eqref{cesc} in \eqref{srenyiesc}, we finally obtain the relation between the entanglement entropy and the effective action on $S^3$
\beq
S= - S_{\mathrm{eff}}^{S^3} + \frac16 \pi \rho (\rho ^2 -1) \cot(\pi \rho)\,. \label{relesc1}
\eeq

From this last equation, the UV limit ($ma \rightarrow 0$ or, equivalently $\rho \rightarrow \frac12$\,) can be trivially taken
\beq \lim_{ma \rightarrow 0} S = - \left.S_{\mathrm{eff}}^{S^3}\right\vert_{ma=0}\,.\nn\eeq

Now, we show that the UV fixed point is not stable, a point that was raised in reference \cite{klestability} and explicitly proved in \cite{dowkerrenyi}. In order to do that, we compute the derivative of $S$ with respect to $\rho$ at $\rho=\frac12$, which is, up to a sign, the derivative with respect to $(ma)^2$ at $ma=0$.
As it is well known, the derivative of $S_{\mathrm{eff}}^{S^3}$ vanishes at this point \cite{kleRenyi}. From equation \eqref{relesc1}, it is easy to see that
\beq \left.\frac{d}{d(ma)^2}S\right\vert_{ma=0}=- \frac{\pi ^2}{16}\,,\eeq
in agreement with \cite{dowkerrenyi}.

Let us now turn to the study of the entanglement entropy near the IR fixed point. To this end, one should notice that even though equation \eqref{relesc1} was obtained near the UV limit (for $(ma)^2 < \frac14 $),  it can be analytically extended to the interval $(ma)^2 \geq \frac14 $. This extension, performed by making the substitution $\rho\rightarrow i \delta$ (with $\delta=\sqrt{(ma)^2 -1/4}$\,) in this equation, gives the result
\beq
S= - S_{\mathrm{eff}}^{S^3} - \frac16 \pi \delta (\delta ^2 +1) \coth(\pi \delta)\,, \label{relesc2}\eeq
and coincides with the result one would obtain by Mellin transforming the zeta function and inverting the heat kernel, as we have already done for the effective action on $S^3$.

\begin{figure}[h]
	\centering
	\begin{minipage}{.4\textwidth}
		\centering
		\includegraphics[height=42mm]{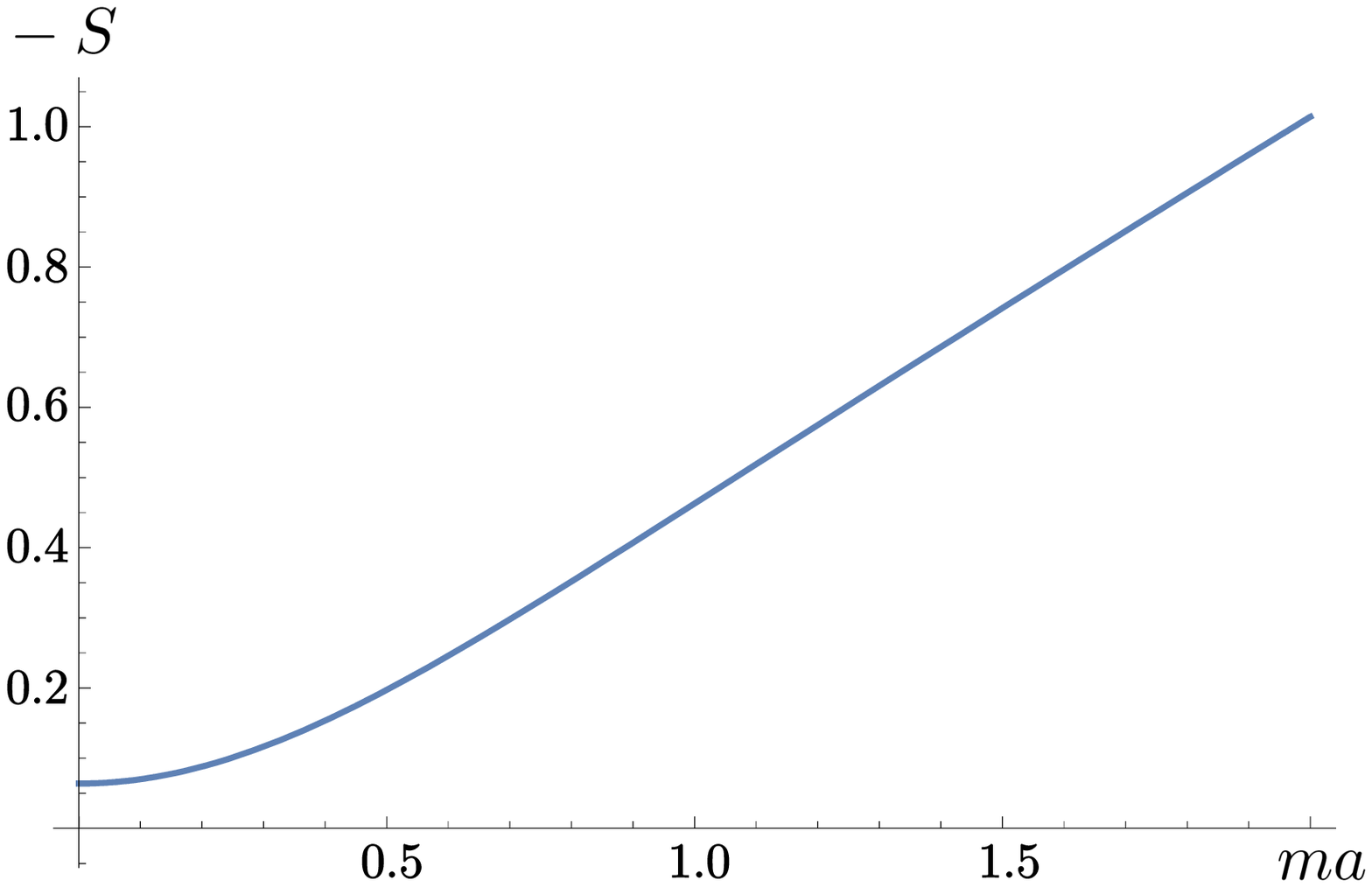}
		\caption{\small Minus the entanglement entropy for a real scalar field as a function of $ma$}
		\label{figure3}
	\end{minipage}
	\hspace{0.02\textwidth}
	\begin{minipage}{.4\textwidth}
		\centering
		\includegraphics[height=42mm]{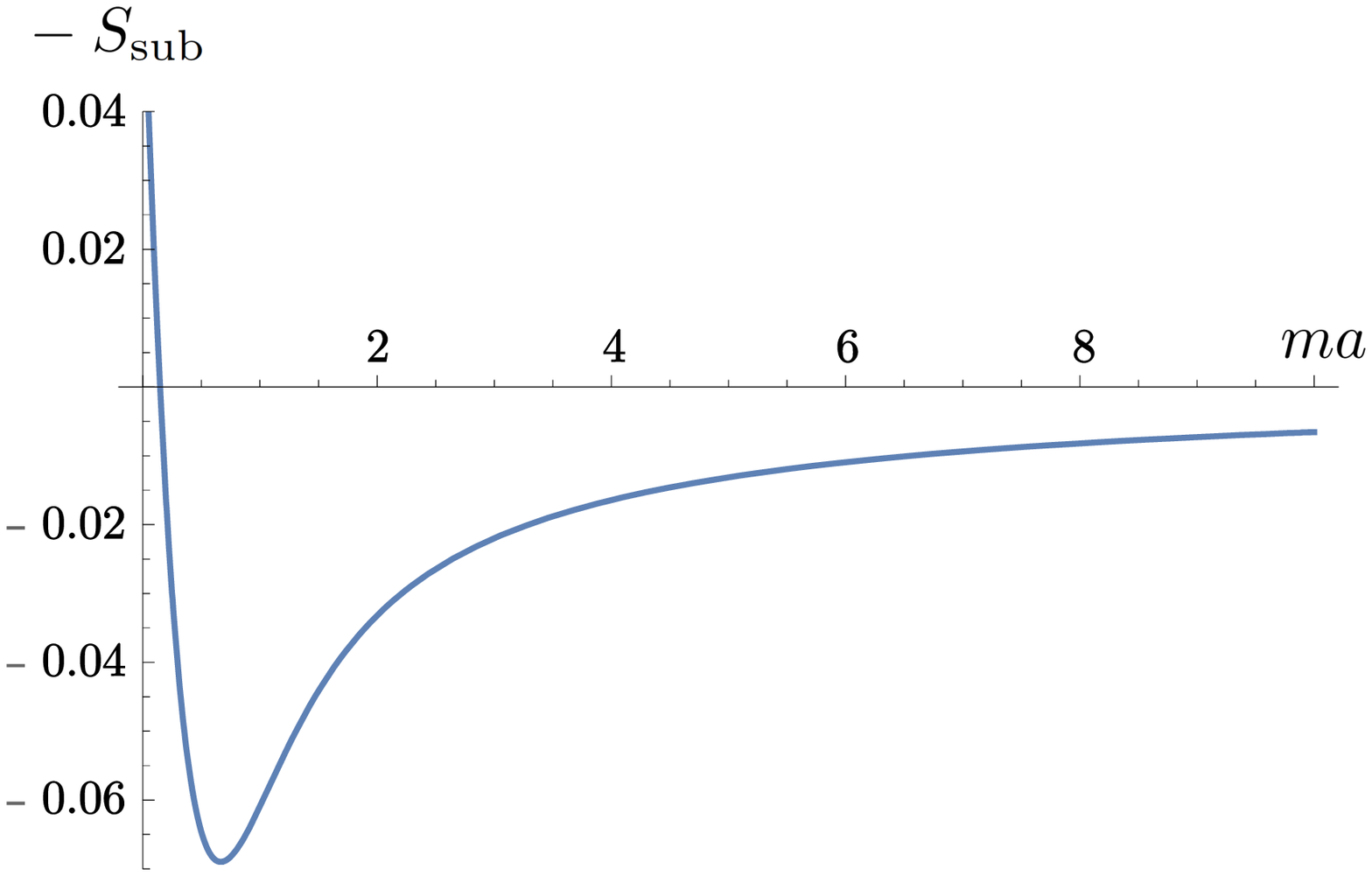}
		\caption{\small Minus the entanglement entropy for a real scalar field where the leading IR divergence has been subtracted}
		\label{figsub}
	\end{minipage}
\end{figure}

Taking into account equations \eqref{relesc2} and \eqref{seffS3}, we see that the entanglement entropy behaves, in the IR limit, as
$-\frac{\pi}{6} ({\delta}^2)^{\frac12}$. As discussed in \cite{dowkerrenyi}, such divergence was to be expected from the asymptotic expansion of the relevant heat kernel. As mentioned in the same reference, in a QFT context one should subtract this divergence, since infinitely massive modes should decouple \cite{dewitt}. As in the case of the effective action, subtracting the whole divergence would spoil unitarity, while subtracting only the leading divergence ($\frac{\pi}{6}\,ma$) to $-S$ gives a not-always-positive nor monotonic function, with a derivative with respect to $(ma)^2$ divergent as $(ma)^{-1}$ at the UV fixed point.
The behavior of minus the IR divergent entanglement entropy for a real scalar field as a function of $ma$ is shown in Figure \ref{figure3}, while the subtracted function appears in Figure \ref{figsub}.

\subsection{Massive Dirac field}

The eigenvalues and degeneracies corresponding to the relevant first order operator in three dimensions, for multiple ($q$) coverings of $S^3$ can be retrieved from the ones in reference \cite{kleRenyi}. In the massive case, they are
\beq
\lambda_n^{\pm}= \pm i\frac{(n+1+\frac{p}{q}+\frac{1}{2q})}{a} + m\,, \qquad n= 0,1,\ldots , \infty \qquad p= 0,1,\ldots , q-1
\eeq
with corresponding degeneracies
\beq
d_n= (n+1)(n+2)\,.
\eeq

The zeta function is, then, given by
\beq
\zeta^{qS^3}(s)=& (a \mu)^{s} & \left\{\sum_{k=0}^{\infty}\sum_{p=0}^{q-1} k(k+1) \left[i\left(k + \frac{p}{q}+\frac{1}{2q}\right)+ma\right]^{-s}\right. \nn \\
&& + \left. \sum_{k=0}^{\infty}\sum_{p=0}^{q-1} k(k+1) \left[-i\left(k + \frac{p}{q}+\frac{1}{2q}\right)+ma\right]^{-s}\right\}\,.
\eeq

After summing and subtracting the needed terms to the degeneracy, this can be cast in the form
\beq
\zeta^{qS^3}(s)=&(a \mu)^{s} & \Bigg\{e^{i\frac{\pi}{2} s}\sum_{p=0}^{q-1} \Bigg[{\zeta}_H \left(s-2,\frac{p}{q}+\frac{1}{2q}+ima\right)\nn\\
  && +\left(1-2\left(\frac{p}{q}+\frac{1}{2q}+ i ma\right)\right){\zeta}_H \left(s-1,\frac{p}{q}+\frac{1}{2q}+ima\right)\nn \\
&& - \left(\frac{p}{q}+\frac{1}{2q}+ i ma\right)\left(1-\left(\frac{p}{q}+\frac{1}{2q}+ i ma\right)\right){\zeta}_H \left(s,\frac{p}{q}+\frac{1}{2q}+ima\right)\Bigg]\nn\\
&& +  e^{-i\frac{\pi}{2} s}\sum_{p=0}^{q-1} \left[ (ma\rightarrow -ma)\right]\Bigg\}
\,.
\label{zetaqdirac}\eeq

As done in the previous cases, and for the same reason (it is easy to check that $\zeta^{qS^3}(0)=0$), we will, from now on, ignore the pre-factor $(a \mu)^{s}$. So, the effective action is given by
\beq
S_{\mathrm{eff}}^{qS^3}=&-&\frac{\pi ma}{12q}\left[1+(2+4(ma)^2)q^2\right]+\sum_{p=0}^{q-1} \left[{\zeta}^{\prime}_H \left(-2,\frac{p}{q}+\frac{1}{2q}+ima\right)+(ma\rightarrow -ma)\right]\nn \\
&+& \sum_{p=0}^{q-1} \left[\left(1-2\left(\frac{p}{q}+\frac{1}{2q}+ i ma\right)\right){\zeta}^{\prime}_H \left(-1,\frac{p}{q}+\frac{1}{2q}+ima\right)+(ma\rightarrow -ma)\right]\nn \\
&-& \sum_{p=0}^{q-1} \Bigg[\left(1-\left(\frac{p}{q}+\frac{1}{2q}+ i ma\right)\right)\left(\frac{p}{q}+\frac{1}{2q}+ i ma\right){\zeta}^{\prime}_H \left(0,\frac{p}{q}+\frac{1}{2q}+ima\right)\nn\\
&+&(ma\rightarrow -ma) \Bigg]\,.
\label{qs3}
\eeq

From the previous expression, it is easy to see that $S_{\mathrm{eff}}^{qS^3}$ reduces (as it should) to $S_{\mathrm{eff}}^{S^3}$, where $S_{\mathrm{eff}}^{S^3}$ is given in equation \eqref{seffdirac} for $q=1$. As a consequence, the R\'enyi entanglement entropy, defined in equation \eqref{Renyi}, will have its first contribution coming from $S_{\mathrm{eff}}^{qS^3}$. To complete the evaluation, one must calculate the derivative of \eqref{qs3} at $q=1$. We do this for the first term, which has its origin in the spectral asymmetry. It contributes to the entanglement entropy with
\beq
S_1=-\frac{\pi ma}{12}\left[1+4(ma)^2\right]\,. \label{1}\eeq

As for the remaining contributions to the R\'enyi entropy, it is convenient to evaluate, first, the derivative with respect to $q$ at $q=1$ of the corresponding terms in zeta function \eqref{zetaqdirac} and, after that, perform the $s$-derivative and take its value at $s=0$. Proceeding this way, the contribution of the first term between square brackets in equation \eqref{qs3} is easily obtained by first using the  summation formula reproduced in \eqref{a21} and then performing the derivatives with respect to $q$ and $s$, making use of the result in \eqref{a22}. After doing so, we determine its contribution to the R\'enyi entropy, which is
\beq
S_2&=&\left[{\zeta}_H \left(-2,\frac{1}{2}+ima\right)-2{\zeta}^{\prime}_H \left(-2,\frac{1}{2}+ima\right)-i ma {\zeta}_H \left(-1,\frac{1}{2}+ima\right)\right.\nn \\&& +\left. 2i ma {\zeta}^{\prime}_H \left(-1,\frac{1}{2}+ima\right)\right]+(ma\rightarrow -ma)\,.\label{2}\eeq

The two remaining sums in \eqref{qs3} are a bit more cumbersome to evaluate. Take for instance
\beq
{\zeta_3}(s)&=&\left\{\sum_{p=0}^{q-1}\left[1-2\left(\frac{p}{q}+\frac{1}{2q}+ i ma\right)\right]{\zeta}_H \left(s-1,\frac{p}{q}+\frac{1}{2q}+ima\right)\right\}+(ma\rightarrow -ma)\nn \\&=&
\Bigg\{\left(1-\frac{1}{q}-2 i ma\right){\zeta}_H \left(s-1,\frac{1}{2q}+ima\right)\nn\\
&&+\sum_{p=1}^{\infty}\left[1-2\left(\frac{p}{q}+\frac{1}{2q}+ i ma\right)\right]{\zeta}_H \left(s-1,\frac{p}{q}+\frac{1}{2q}+ima\right)\nn \\
&& -\sum_{p=q}^{\infty}\left[1-2\left(\frac{p}{q}+\frac{1}{2q}+ i ma\right)\right]{\zeta}_H \left(s-1,\frac{p}{q}+\frac{1}{2q}+ima\right)\Bigg\}+(ma\rightarrow -ma)\nn\\
&=&\Bigg\{\left(1-\frac{1}{q}-2 i ma\right){\zeta}_H \left(s-1,\frac{1}{2q}+ima\right)\nn\\
&&+\sum_{p=1}^{\infty}\left[1-2\left(\frac{p}{q}+\frac{1}{2q}+ i ma\right)\right]{\zeta}_H \left(s-1,\frac{p}{q}+\frac{1}{2q}+ima\right)\nn \\
&& -\sum_{p=0}^{\infty}\left[1-2\left(\frac{p}{q}+\frac{1}{2q}+ i ma\right)-2\right]{\zeta}_H \left(s-1,\frac{p}{q}+\frac{1}{2q}+i ma+1\right)\Bigg\}+(ma\rightarrow -ma)\nn\,.\eeq

From this expression, after shifting the index in the last sum, the derivative with respect to $q$ can be performed, and evaluated at $q=1$ to obtain
\beq \left.\frac{d}{dq}\right\vert_{q=1}{\zeta_3}(s)&=&2\Bigg\{\frac12 {\zeta}_H \left(s-1,\frac{1}{2}+ima\right)-\frac{(s-1)i ma}{2}{\zeta}_H \left(s,\frac{1}{2}+ima\right)\nn\\
&&+\sum_{p=0}^{\infty}{\zeta}_H \left(s-1,p+\frac{3}{2}+ima\right)\nn \\ \nn
&& -(s-1)\sum_{p=0}^{\infty}(2p+2+2i ma){\zeta}_H \left(s,p+\frac{3}{2}+ima\right)\Bigg\}+(ma\rightarrow -ma)\,.\eeq
At this point, the sums over $p$ can be calculated by using our results \eqref{a25} and \eqref{a26}. Finally, the $s$ derivative at $s=0$ can be computed. After doing so we get for the corresponding contribution to the R\'enyi entanglement entropy
\beq
S_3&=&\left\{3{\zeta}^{\prime}_H \left(-2,\frac{1}{2}+ima\right)-2\left(\frac12+i ma\right) {\zeta}^{\prime}_H \left(-1,\frac{1}{2}+ima\right)-{\zeta}_H \left(-2,\frac{1}{2}+ima\right)\right.\nn  \\
&& -\left(\frac12+i ma\right)^2 {\zeta}^{\prime}_H \left(0,\frac{1}{2}+ima\right)+{\left(\frac12+i ma\right)}^2 {\zeta}_H \left(0,\frac{1}{2}+ima\right)+{\zeta}^{\prime}_H \left(-1,\frac{1}{2}+ima\right)\nn  \\
&& +\left. i ma \left[{\zeta}^{\prime}_H \left(0,\frac{1}{2}+ima\right)-{\zeta}_H \left(0,\frac{1}{2}+ima\right)\right]\right\}+(ma\rightarrow -ma)\,.\label{3}\eeq

The contribution from the third sum in \eqref{qs3}  can be obtained in the same manner, this time using, also, our result \eqref{a27}. This gives
\beq S_4=\frac{1}{24} &&\!\!\!\! \left\{8{\zeta}_H \left(-2,\frac{1}{2}+ima\right)-8{\zeta}_H \left(0,\frac{1}{2}+ima\right)-24 i ma\, {\zeta}^{\prime}_H \left(0,\frac{1}{2}+ima\right)\right.\nn \\
&& -24{\zeta}^{\prime}_H \left(-2,\frac{1}{2}+ima\right)+\left[6+24i ma-24(ma)^2\right]{\zeta}^{\prime}_H \left(0,\frac{1}{2}+ima\right)\nn \\
&& -\left.\left[8i (ma)^3 +2ima\right]\psi\left(\frac12+i ma\right)\right\}+(ma\rightarrow -ma)\,, \label{4}\eeq
where $\psi(x)$ is the Polygamma function.

\begin{figure}[h]
\centering
\includegraphics[height=42mm]{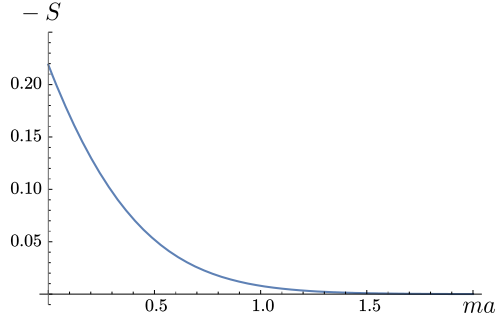}
\caption{\small Minus the entanglement entropy for the Dirac field as a function of $ma$}
\label{figure4}
\end{figure}

The sum of $-S_{\mathrm{eff}}^{S^3}$ and the contributions \eqref{1} to \eqref{4} gives an analytic expression for the entanglement entropy, as defined in equation \eqref{Renyi}, which is
\beq
S=-S_{\mathrm{eff}}^{S^3}+\frac{\pi ma}{12}\left[1+4(ma)^2\right]\left[\tanh{(\pi ma)}-1\right]\,,
\label{renyidirac}\eeq
where $S_{\mathrm{eff}}^{S^3}$ is given by equation \eqref{seffdirac}. As expected, minus the entanglement entropy coincides with the effective action on the round sphere at the UV fixed point. Moreover, it vanishes in the IR limit. It is always positive and decreases to zero under the flow of the RG. It fails to be stable in the UV, where its derivative with respect to $ma$ equals $-\frac{\pi}{6}$. Such behavior is shown in Figure \ref{figure4}.

\section{Renormalized entanglement entropy}\label{sect4}

Motivated by the proposals made in references \cite{casini,liu} we analyze, in this section, the behavior of the so-called renormalized entanglement entropy, in order to determine whether it can be considered a good candidate for a $\textit{C}$-function in three dimensions. Its definition is
\beq
S_{\mathrm{ren}}=ma\,\frac{d S}{d(ma)}-S\,,
\eeq
where $S$ is the entanglement entropy studied in the previous section. In this section, we show that this is an ideal $\textit{C}$-function for free Dirac fields, while it is far from being so in the case of a real scalar field. We start by analyzing the encouraging case of the Dirac field.

\subsection{Massive Dirac field}

Making use of the explicit results in equations \eqref{seffdirac} and \eqref{renyidirac}, a quite direct calculation leads to
\beq
ma\,\frac{d S}{d(ma)}=\frac{\pi\,ma}{6}\left[1-\tanh{(\pi\,ma)}\right]+\frac{\pi\,(ma)^2}{12}\frac{\left[1+4(ma)^2\right]}{\cosh^2{(\pi\,ma})}\,.
\eeq

Thus,
\beq
S_{\mathrm{ren}}=S_{\mathrm{eff}}^{S^3}+\frac{\pi\,ma}{12}\left[3+4(ma)^2\right]\left[1-\tanh{(\pi\,ma)}\right]+\frac{\pi\,(ma)^2}{12}\frac{\left[1+4(ma)^2\right]}{\cosh^2{(\pi\,ma})}\,.
\eeq

\begin{figure}[h]
\centering
\includegraphics[height=42mm]{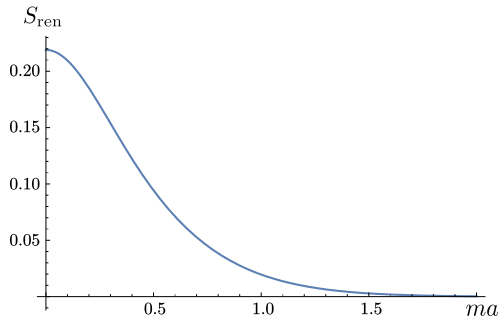}
\caption{\small Renormalized entanglement entropy for the Dirac field as a function of $ma$}
\label{figure5}
\end{figure}

Now, it is easy to verify that this function is, indeed, a monotonic positive function, vanishing in the IR limit and stable, i.e., with a vanishing derivative with respect to $ma$ at the UV fixed point. Moreover, its value at this fixed point coincides with the value of the effective action on $S^3$, as expected \cite{casini}. It is depicted in Figure \ref{figure5}.

\subsection{Massive scalar field}

Things are entirely different when it comes to the free massive scalar field. Starting from the expressions valid for $(ma)^2<\frac14$, the renormalized entanglement entropy can be written as
\beq
S_{\mathrm{ren}}=-S+\frac{(ma)^2}{\rho}\left(\frac{\pi}{6}\cot{(\pi\,\rho)}+\frac{{\pi}^2}{6}({\rho}^3 -\rho)\csc^2{(\pi \rho)}\right)\,,
\eeq
where as before $\rho=\sqrt{1/4-(ma)^2}$\,, and $S$ is the entanglement entropy given in equation \eqref{relesc1}.

This expression can be analytically continued to the whole range of $ma$ and coincides with the expression obtained directly from equation \ref{relesc2}.

\begin{figure}[h]
\centering
\includegraphics[height=42mm]{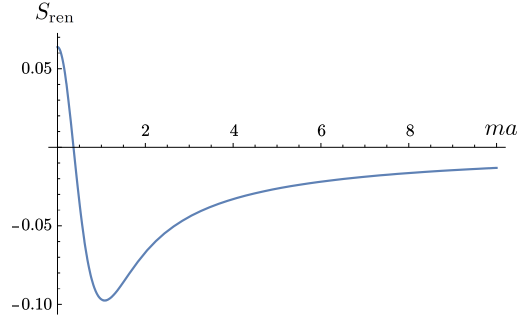}
\caption{\small Renormalized entanglement entropy for the real scalar field as a function of $ma$}
\label{figure6}
\end{figure}

It is obvious that, at $ma=0$, its value coincides with the value of the effective action on $S^3$. It is also true that this definition of the renormalized entanglement entropy does eliminate the IR divergence. However, as we can see in Figure \ref{figure6}, this gives a function which goes through negative values under the flow of the renormalization group and, therefore, it is clearly ruled out as a measure of degrees of freedom.

A more ``natural'' definition would be, in the scalar case, a function we call the modified renormalized entanglement entropy,
\beq
{\tilde{S}}_{\mathrm{ren}}=\rho\,\frac{d S}{d\rho}-S\,.
\eeq

\begin{figure}[h]
\centering
\includegraphics[height=42mm]{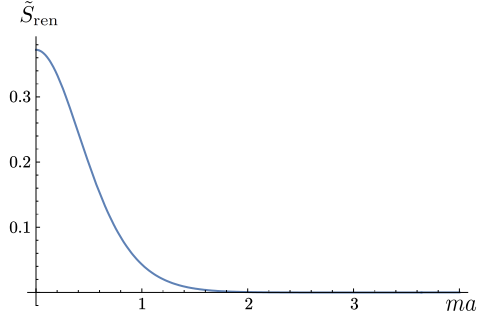}
\caption{Modified renormalized entanglement entropy for the real scalar field as a function of $ma$}
\label{figure7}
\end{figure}

This definition would lead to
\beq
{\tilde{S}}_{\mathrm{ren}}=S_{\mathrm{eff}}^{S^3}-\frac{{\pi}^2}{6}({\rho}^4 -{\rho}^2)\csc^2{(\pi \rho)}-\frac{\pi}{6}{\rho}^3\cot{(\pi\,\rho)}\,,\eeq
for $(ma)^2<\frac14$, and to its evident analytical extension for $(ma)^2>\frac14$. The behavior of this function with $ma$ is shown in Figure \ref{figure7}.

While being always positive and decreasing under the flow of the renormalization group, this new function is not stable at the massless fixed point. In fact, the value of its derivative at this point is $\frac{-5{\pi}^2}{48}$. Moreover, the value of the function itself at the UV fixed point ($\rho=\frac12$) differs from the value of the effective action at the same point by $\frac{{\pi}^2}{32}$. As a consequence, no $F$-theorem guarantees its positive character at this fixed point for other conformal theories.

\section {A new candidate}\label{sect5}

Let us summarize what we have up to this point: we have performed a fully analytic study on the three-sphere of the main candidates for $C$-functions, as applied to free massive scalar and Dirac fields. While all of them are, indeed, good candidates for $C$-functions in the case of the Dirac field, none of them is so in the scalar case. In fact, our analytic calculations show that,
\begin{itemize}
  \item The effective action on $S^3$, while satisfying requirement \ref{r1}, being monotonically decreasing (though not UV stable) and vanishing in the IR limit for Dirac fields, presents, in the scalar case, an IR divergence which cannot be renormalized without breaking unitarity or getting a not-everywhere positive nor monotonic function, with a divergence in the derivative with respect to the coupling constant at the UV fixed point.
    \item The R\'enyi entanglement entropy, computed via the replica trick has, for Dirac fields, the same characteristics as the effective action on the sphere, which would also make it a candidate for a $C$-function, since it satisfies the requirements \ref{r1} and \ref{r2}. Unfortunately, as noted before \cite{dowkerrenyi}, an IR divergence due to the conical singularity appears in the scalar case. While being milder than in the previous case, this divergence is equally difficult to renormalize.
      \item Finally, the renormalized entanglement entropy satisfies, in the Dirac case, all the requirements demanded from a $C$-function, including the stability at the UV fixed point. However, when evaluated for the scalar field, one finds a function that, while satisfying $C_{UV}>C_{IR}$ and vanishing in the IR limit, not only is non-monotonic, but it also becomes negative for part of its evolution under the RG flow. A more satisfactory modified renormalized R\'enyi entanglement entropy can be defined. However, this function, in addition to being unstable at the UV fixed point, fails to coincide with the effective action at the same point. So, no $F$-theorem would guarantee its positive character at such fixed point for other theories.
\end{itemize}

In this section, we present a new candidate for a $C$-function for free theories on $S^3$, which we call modified $F$-function.

\subsection{Massive Dirac field}

As already mentioned, in the case of the Dirac field, when the decoupling phase of the determinant is chosen, the effective action does not need an IR renormalization. Still, it is clearly unstable at the UV fixed point. In fact, its derivative with respect to $ma$ differs from zero at $ma=0$. The same is true for minus the entanglement entropy in equation \eqref{renyidirac}. From equations \eqref{derdirac} and \eqref{renyidirac}, it is easy to check that, in this case, $-S=S_{\mathrm{eff}}-\frac13 ma\,\frac{d}{d(ma)}S_{\mathrm{eff}}$. Motivated by this remark, we define a modified effective action,
\beq
\tilde{F}^D=S_{\mathrm{eff}}-\frac13 ma\,\frac{d}{d(ma)}S_{\mathrm{eff}}=S_{\mathrm{eff}}+\frac{\pi}{3} ma \left[\frac14 +(ma)^2\right]\left[1-\tanh(\pi m a)\right]\,.\eeq

Taking into account that, for this theory, the coupling constant is $g^{D}=ma$, one has
\beq
{\tilde{F}^D}= S_{\mathrm{eff}}-\frac{\omega}{d} g^{D}\,\frac{d\,S_{\mathrm{eff}}}{d(g^{D})}\,,\label{fmod}\eeq
where $\omega=1$ is the order of the Dirac operator and $d=3$ is the dimension of the manifold $S^3$.

\subsection{Massive scalar field}

For the scalar theory, things are more complicated than in the Dirac case. Among all the possibilities that we have already explored, there is not a single function nonnegative, monotonic, vanishing at the IR limit and coinciding, at the UV point, with the $F$-function. The analog of equation \eqref{fmod} is, in this case:
\beq
{\tilde{F}}^{S}=S_{\mathrm{eff}}-\frac13 \delta\,\frac{d}{d(\delta)}S_{\mathrm{eff}}=S_{\mathrm{eff}}+\frac{\pi}{6}{\delta}^3 \coth(\pi \delta)\,,\eeq
where, as before, ${\delta}^2=(ma)^2 -\frac14$.
This can also be written as
\beq
\tilde{F}^{S}=S_{\mathrm{eff}}-\frac{\omega}{d} g^{S}\,\frac{d\,S_{\mathrm{eff}}}{d(g^{S})}\,,\eeq
where we have taken $g^{S}=(ma)^2 -\frac14$, which is the naturally appearing dimensionless coupling constant, as can be seen by comparing the IR divergence of the effective action on $S^3$ with the corresponding one on $R^3$, which equals ${(ma)}^3$. Moreover, $\omega=2$ is the order of the relevant, Laplacian-type operator and, again, $d=3$ is the dimension of the manifold.

\begin{figure}[h]
	\centering
	\begin{minipage}{.4\textwidth}
		\centering
		\includegraphics[height=42mm]{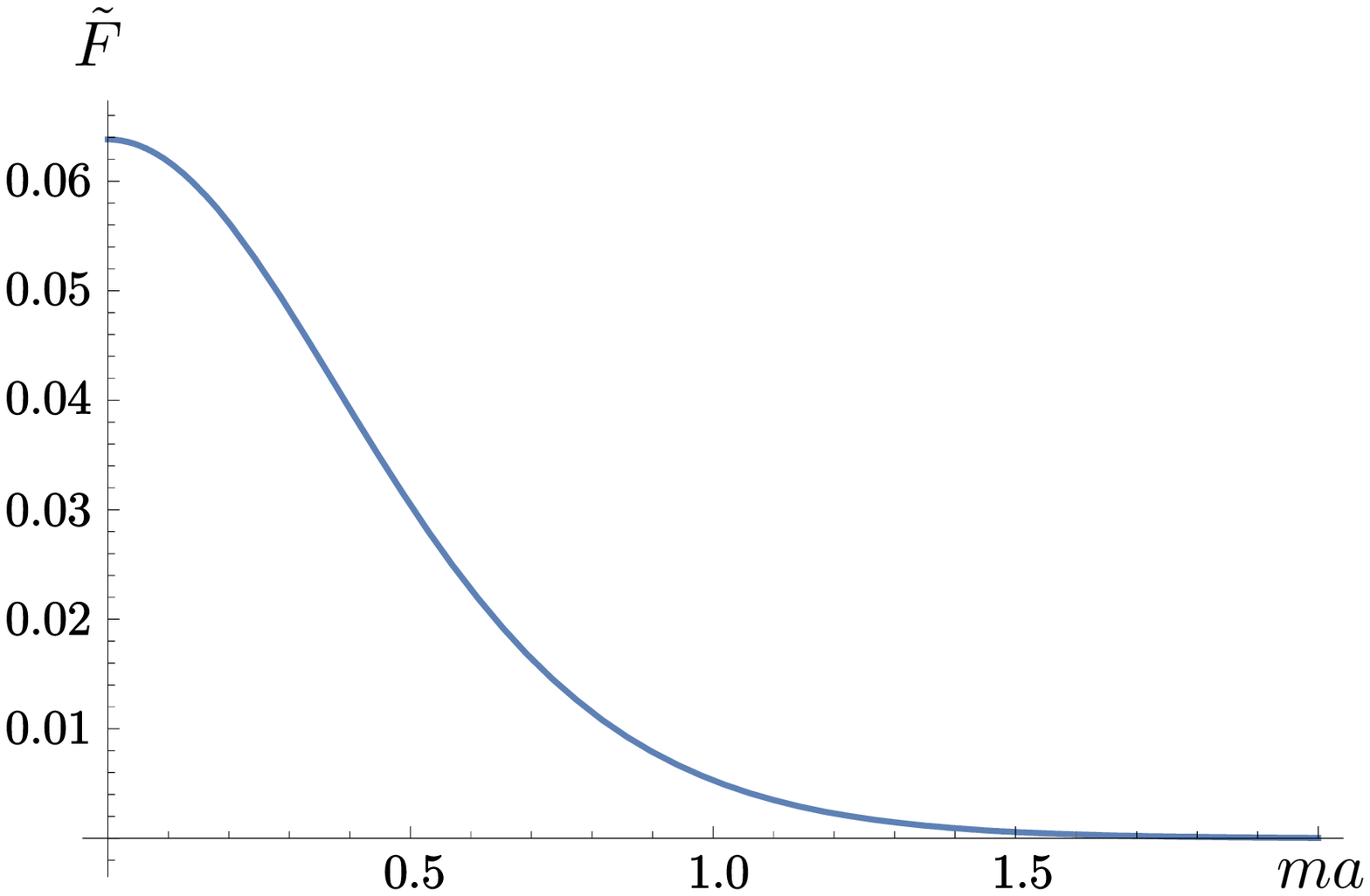}
		\caption{\small Modified $F$-function for the free scalar field as a function of $ma$}
		\label{figure10}
	\end{minipage}
	\hspace{0.02\textwidth}
	\begin{minipage}{.4\textwidth}
		\centering
		\includegraphics[height=42mm]{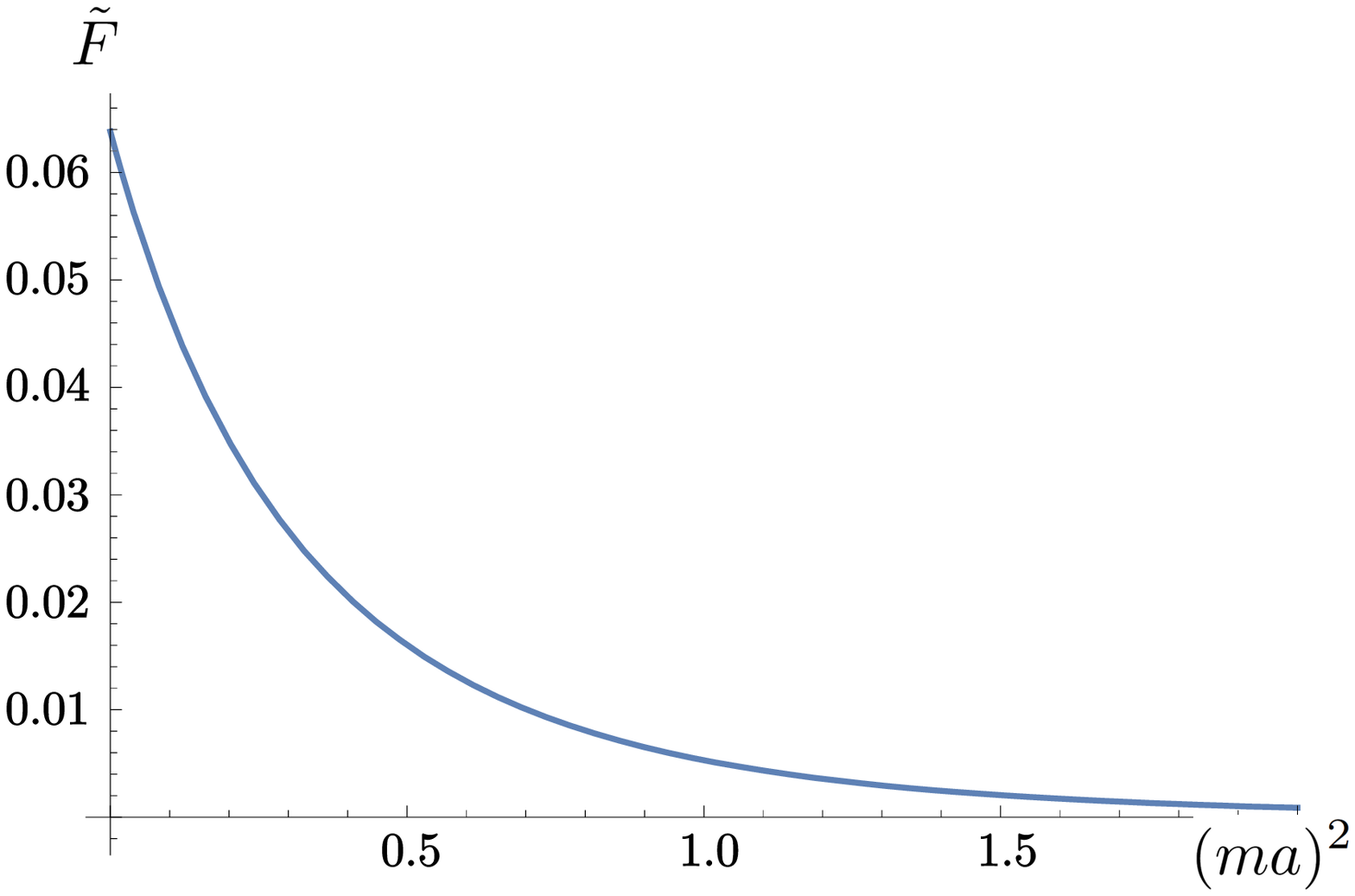}
		\caption{\small Modified $F$-function for the free scalar field as a function of ${(ma)}^2$}
		\label{figure11}
	\end{minipage}
\end{figure}

It is evident that this new candidate coincides with $F$ at the UV fixed point. It is also clear that it satisfies the decoupling condition in the IR. Moreover, it can be shown to be a positive monotonic function, vanishing in the IR limit. Its behavior as a function of $ma$ appears in Figure \ref{figure10}.

As in the Dirac case, it fails to be UV-stable, as it can be seen in Figure \ref{figure11}.

\section {Conclusions}

In a previous paper \cite{Asorey:2014gsa} we analyzed a $C$-function proposal for theories on other three-dimensional manifolds, i.e., lens spaces. Such functions are built from a topological term in the high-temperature expansion of the entropy. Further work on this subject is in progress.

Here, we have presented a detailed analytic study on the three-sphere of some popular candidates for $C$-functions, both for scalar and Dirac free massive theories. In the Dirac case, all these functions (effective action, R\'enyi entanglement entropy and renormalized entanglement entropy) can be considered good $C$-functions, once the decoupling phase has been selected in the evaluation of the Dirac determinant. Among them, the renormalized entanglement entropy is particularly interesting since it is the only UV-stable one. On the contrary, due to the presence of IR divergencies, neither the effective action nor the entanglement entropy can be considered $C$-functions for the scalar theory once their IR divergencies have been subtracted. As we have also shown, the so-called renormalized entanglement entropy isn't a good candidate either, since it fails to be always positive. A variant of this last one, which we have called the modified renormalized entanglement entropy, is also not satisfactory, since it fails to coincide with $F$ at the UV fixed point.

In view of these negative results, we introduced a new proposal, which we call modified $F$-function, defined by subtracting from the effective action the quotient of the order of the operator and the dimension of the manifold times its logarithmic derivative with respect to the relevant dimensionless coupling constant. This new function is, for the very simple theories considered here, always positive, monotonously decreasing towards the IR, where it vanishes, and coincides at the UV fixed point with the effective action or $F$-function. It only fails to satisfy the UV-stability condition \ref{r3}. This is more than what any of the other candidates can claim, since none of them is satisfactory both for free scalar and Dirac fields. Still, it is not clear for us to which extent its definition, in terms of the logarithmic derivative with respect to a ``natural coupling constant'', can be extended to interacting theories on $S^3$ or even to noninteracting ones on higher-dimensional odd spheres. Additionally, some new results concerning the evaluation of series of zeta functions, which were needed along our calculations are included in the appendix.

\section*{Acknowledgments}
{The authors thank Manuel Asorey for many useful discussions and comments. Research of CGB, DD and EMS has been partially supported by CONICET PIP681 and UNLP Proyecto 11/X748 (Argentina).
The work of ICP has been partially supported by the Spanish MINECO/FEDER grant FPA2015-65745-P and DGA-FSE grant 2015-E24/2.}

\appendix
\section{Some useful identities involving Hurwitz' zeta functions}\label{ap2}

We begin this appendix by listing two well-known results involving Hurwitz' zeta functions. The first one is the so-called multiplication formula,
\beq
\sum_{p=0}^{q-1} \zeta_H (s, \frac pq +z) = q^s \zeta_H (s, qz)\,, \qquad q \in \mathbb{N} \,.\label{a21}
\eeq

Also well-known are
\beq \frac{d}{dx} \zeta_H (s,x) = -s \zeta_H (s+1,x)\,,
\label{a22}
\eeq
\beq \frac{d}{ds}\frac{d}{dx} \zeta_H (s,x) = -\zeta_H (s+1,x) - s \zeta^{\prime}_H (s+1,x)\,,
\label{a23}
\eeq
and
\beq \zeta_H (s,x+1) =  \zeta_H (s,x) - x^{-s} \,.
\label{a24}
\eeq
Now, another known result (see, for instance, reference \cite{sri}) is
\beq
\sum_{p=0}^{\infty}  \zeta_H (s,p+x)= \zeta_H (s-1,x-1) - (x-1) \zeta_H (s, x-1)\,. \label{a25}
\eeq

In what follows, we derive other useful formulas. We start with
\beq
\sum_{p=0}^{\infty}  p \, \zeta_H (s,p+x) &=& \frac12 \left[\zeta_H (s-2,x-1) - (x-1) \zeta_H (s-1, x-1)\right] \nn \\
&-&\frac x2 \left[\zeta_H (s-1,x-1) - (x-1) \zeta_H (s, x-1)\right] \,.\label{a26}\eeq

To obtain the previous expression, we have used the Mellin transform to write the zeta function
\beq
\sum_{p=0}^{\infty}  p \, \zeta_H (s,p+x) &=& \sum_{p=0}^{\infty}  p \sum_{n=0}^{\infty} \frac{1}{\Gamma(s)}\int_0^{\infty}dt\,t^{s-1}e^{-(n+p+x)t}\nn\\
&=&\frac{1}{\Gamma(s)}\sum_{p=0}^{\infty}\int_0^{\infty}dt\,t^{s-1} \frac{e^{-xt}}{1-e^{-t}}\left(-\frac{d}{dt}(e^{-pt})\right)\nn\\
&=&\frac{1}{\Gamma(s)}\int_0^{\infty}dt\,t^{s-1} \frac{e^{-xt}}{1-e^{-t}}\left(-\frac{d}{dt}\left(\frac{1}{1-e^{-t}}\right)\right)\nn\\
&=&\frac{1}{2} \frac{1}{\Gamma(s)}\int_0^{\infty}dt\,t^{s-1} {e^{-xt}}\left(-\frac{d}{dt}\left(\frac{1}{1-e^{-t}}\right)^2 \right)\nn\\
&=&\frac{1}{2} \frac{(s-1)}{\Gamma(s)}\int_0^{\infty}dt\,t^{s-2} \frac{e^{-xt}}{(1-e^{-t})^2} -
\frac{1}{2} \frac{1}{\Gamma(s)}\int_0^{\infty}dt\,t^{s-1}\frac{x\, e^{-xt}}{(1-e^{-t})^2}\,,\nn
\eeq
where we have integrated by parts and used the fact that for $\Re(s)>3$ the integrated term vanishes. Once again, we write $\frac{1}{(1-e^{-t})^2}=-e^t\frac{d}{dt}\left(\frac{1}{1-e^{-t}}\right)$ and perform the corresponding integrations by parts to obtain
\beq
\sum_{p=0}^{\infty}  p \, \zeta_H (s,p+x) &=&\frac{1}{2 \Gamma(s-1)}\left[(s-2)\int_0^{\infty}dt\,t^{(s-2)-1}\, \frac{e^{-(x-1)t}}{1-e^{-t}}\right.\nn\\
&&\left.\qquad\qquad\qquad- (x-1)\int_0^{\infty}dt\,t^{(s-1)-1}\, \frac{e^{-(x-1)t}}{1-e^{-t}}\right]\nn\\
&-&\frac{x}{2 \Gamma(s)}\left[(s-1)\int_0^{\infty}dt\,t^{(s-1)-1}\, \frac{e^{-(x-1)t}}{1-e^{-t}}- (x-1)\int_0^{\infty}dt\,t^{s-1} \,\frac{e^{-(x-1)t}}{1-e^{-t}}\right]\nn\\
&=&\frac12 \left[\zeta_H (s-2,x-1) - (x-1) \zeta_H (s-1, x-1)\right] \nn \\
&-&\frac x2 \left[\zeta_H (s-1,x-1) - (x-1) \zeta_H (s, x-1)\right]\nn
\eeq

 Along the same steps, it is easy to show that
\beq
\sum_{p=0}^{\infty}  p^2 \zeta_H (s,p+x) &=&
\frac12 [\zeta_H (s-2,x-1)-(x-1)\zeta_H (s-1,x-1)]\nn\\&-&\frac{x}{2} [\zeta_H (s-1,x-1)-(x-1)\zeta_H (s,x-1)]\nn \\
&+&\frac23\left\{\frac12 [\zeta_H (s-3,x-1)-(x-1)\zeta_H (s-2,x-1)]\right. \nn \\&-&\left.\frac{x}{2} [\zeta_H (s-2,x-1)-(x-1)\zeta_H (s-1,x-1)]\right\}\nn \\
&-&\frac23(x+1)\left\{\frac12 [\zeta_H (s-2,x-1)-(x-1)\zeta_H (s-1,x-1)]\right. \nn \\&-&\left.\frac{x}{2} [\zeta_H (s-1,x-1)-(x-1)\zeta_H (s,x-1)]\right\}\,.\label{a27}
\eeq


\begin{thebibliography}{99}

\bibitem{wilson}
K.G. Wilson and J. Kogut,
\textit{The renormalization group and the $\epsilon$ expansion},
Physics Reports {\bf 12}, 75 (1974).

\bibitem{stueckelberg_petermann}
E.~C.~G.~Stueckelberg and A.~Petermann,
Helv.~Phys.~Acta 26 (1953) 499.

\bibitem{zam}
A.~B.~Zamolodchikov,
\textit{Irreversibility of the Flux of the Renormalization Group in a 2D Field Theory,}
JETP Lett.\  {\bf 43} (1986) 730
[Pisma Zh.\ Eksp.\ Teor.\ Fiz.\  {\bf 43} (1986) 565].

\bibitem{cardy}
P. Calabrese and J. Cardy,
\textit{Entanglement entropy and quantum field theory}
Journal of Statistical Mechanics: Theory and Experiment {\bf 2004}, P06002 (2004).

\bibitem{cardy4}
J.~L.~Cardy,
\textit{Is There a c Theorem in Four-Dimensions?,}
Phys.\ Lett.\ B {\bf 215} (1988) 749.

\bibitem{osborn}
H.~Osborn,
\textit{Derivation of a Four-dimensional $c$ Theorem,}
Phys.\ Lett.\ B {\bf 222} (1989) 97.

\bibitem{osborn-jack}
I.~Jack and H.~Osborn,
\textit{Analogs for the $c$ Theorem for Four-dimensional Renormalizable Field Theories,}
Nucl.\ Phys.\ B {\bf 343} (1990) 647.

\bibitem{KS}
Z.~Komargodski and A.~Schwimmer,
\textit{On Renormalization Group Flows in Four Dimensions,}
JHEP {\bf 1112} (2011) 099
\href{http://arxiv.org/abs/1107.3987}{[arXiv:1107.3987 [hep-th]]}.

\bibitem{K}
Z.~Komargodski,
\textit{The Constraints of Conformal Symmetry on RG Flows,}
JHEP {\bf 1207} (2012) 069
\href{http://arxiv.org/abs/1112.4538}{[arXiv:1112.4538 [hep-th]]}.

\bibitem{Cappelli:1990yc}
A.~Cappelli, D.~Friedan and J.~I.~Latorre,
\textit{C theorem and spectral representation,}
Nucl.\ Phys.\ B {\bf 352} (1991) 616.

\bibitem{Anselmi:1997am}
D.~Anselmi, D.~Z.~Freedman, M.~T.~Grisaru and A.~A.~Johansen,
\textit{Nonperturbative formulas for central functions of supersymmetric gauge theories,}
Nucl.\ Phys.\ B {\bf 526} (1998) 543
\href{http://arxiv.org/pdf/hep-th/9708042.pdf}{[hep-th/9708042]}.

\bibitem{Barnes2004}
E.~Barnes, K.~A.~Intriligator, B.~Wecht, and J.~Wright,
Nucl.\ Phys.\ B{\bf 702} (2004) 131
\href{http://arxiv.org/pdf/hep-th/0408156v2}{[hep-th/0408156]}

\bibitem{Jafferis:2011zi}
D.~L.~Jafferis, I.~R.~Klebanov, S.~S.~Pufu and B.~R.~Safdi,
\textit{Towards the F-Theorem: N=2 Field Theories on the Three-Sphere,}
JHEP {\bf 1106} (2011) 102
\href{http://arxiv.org/abs/arXiv:1103.1181}{[arXiv:1103.1181 [hep-th]]}.

\bibitem{kleF}
I.~R.~Klebanov, S.~S.~Pufu and B.~R.~Safdi,
\textit{F-Theorem without Supersymmetry,}
JHEP {\bf 1110} (2011) 038
\href{http://arxiv.org/abs/arXiv:1105.4598}{[arXiv:1105.4598 [hep-th]]}.

\bibitem{Casini:2011kv}
H.~Casini, M.~Huerta and R.~C.~Myers,
\textit{Towards a derivation of holographic entanglement entropy,}
JHEP {\bf 1105} (2011) 036
\href{http://arxiv.org/abs/arXiv:1102.0440}{[arXiv:1102.0440 [hep-th]]}.

\bibitem{casini}
H.~Casini and M.~Huerta,
\textit{On the RG running of the entanglement entropy of a circle,}
Phys.\ Rev.\ D {\bf 85}, 125016 (2012)
\href{http://arxiv.org/abs/arXiv:1202.5650}{[arXiv:1202.5650 [hep-th]]}.

\bibitem{dowkermasssphere}
J.~S.~Dowker,
\textit{Massive sphere determinants,}
\href{http://arxiv.org/abs/arXiv:1404.0986}{[arXiv:1404.0986 [hep-th]]}.

\bibitem{komleshouches}
Z.~Komargodski,
\textit{Aspects of Renormalization Group Flows,}
in \textit{Theoretical physics to face the challenge of LHC, Lecture Notes of the Les Houches Summer School, Session XCVII}, Oxford university Press (2015).

\bibitem{sri}
H. M. Srivastava, J. Choi,
\textit{Zeta and q-Zeta Functions and Associated Series and Integrals,}
Elsevier (2012).

\bibitem{jackiw}
S. Deser, L. Griguolo and D. Seminara,
\textit{Gauge invariance, finite temperature, and parity anomaly in D= 3},
Phys. Rev. Lett. {\bf 79}, 1976 (1997)
\href{http://arxiv.org/abs/hep-th/9705052}{[hep-th/9705052]}

\bibitem{elizalde}
E.~Elizalde,
\textit{A Simple recurrence for the higher derivatives of the Hurwitz zeta function,}
J.\ Math.\ Phys.\  {\bf 34}, 3222 (1993).

\bibitem{kleRenyi}
I.~R.~Klebanov, S.~S.~Pufu, S.~Sachdev and B.~R.~Safdi,
\textit{Renyi Entropies for Free Field Theories,}
JHEP {\bf 1204}, 074 (2012)
\href{http://arxiv.org/abs/arXiv:1111.6290}{[arXiv:1111.6290 [hep-th]]}.

\bibitem{dowkerentang}
J.~S.~Dowker,
\textit{Entanglement entropy for odd spheres},
\href{http://arxiv.org/abs/arXiv:1012.1548}{[arXiv:1012.1548 [hep-th]]}.

\bibitem{klestability}
 I.~R.~Klebanov, T.~Nishioka, S.~S.~Pufu and B.~R.~Safdi,
\textit{Is Renormalized Entanglement Entropy Stationary at RG Fixed Points?,}
JHEP {\bf 1210} (2012) 058
\href{http://arxiv.org/abs/arXiv:1207.3360}{[arXiv:1207.3360 [hep-th]]}.

\bibitem{dowkerrenyi}
J.~S.~Dowker,
\textit{Sphere Renyi entropies,}
J.\ Phys.\ A {\bf 46} (2013) 225401
\href{http://arxiv.org/abs/arXiv:1212.2098}{[arXiv:1212.2098 [hep-th]]}.

\bibitem{dewitt}
DeWitt, B.S.
\textit{Quantum gravity: the new synthesis in General Relativity}, edited
by S. W. Hawking and W. Israel, CUP, Cambridge, 1979.

\bibitem{dW} B. S. DeWitt,
\textit{Quantum Field Theory in Curved Spacetime},
Phys. Rep. {\bf 19} (1975) 295.

\bibitem{liu}
H.~Liu and M.~Mezei,
\textit{A Refinement of entanglement entropy and the number of degrees of freedom,}
JHEP {\bf 1304} (2013) 162
\href{http://arxiv.org/abs/arXiv:1202.2070}{[arXiv:1202.2070 [hep-th]]}.

\bibitem{Asorey:2014gsa}
M.~Asorey, C.~G.~Beneventano, I.~Cavero-Pel\'{a}ez, D.~D'Ascanio and E.~M.~Santangelo,
\textit{Topological Entropy and Renormalization Group flow in 3-dimensional spherical spaces,}
JHEP {\bf 1501} (2015) 078
\href{http://arxiv.org/abs/arXiv:1406.6602}{[arXiv:1406.6602 [hep-th]]}.

\end{thebibliography}
\end{document}